\documentstyle[amsmath,amssymb,graphicx]{article}

\renewcommand{\arraystretch}{1.3}

\catcode`\@=11
\def\marginnote#1{}

\newcount\hour
\newcount\minute
\newtoks\amorpm
\hour=\time\divide\hour by60
\minute=\time{\multiply\hour by60 \global\advance\minute by-\hour}
\edef\standardtime{{\ifnum\hour<12 \global\amorpm={am}%
        \else\global\amorpm={pm}\advance\hour by-12 \fi
        \ifnum\hour=0 \hour=12 \fi
        \number\hour:\ifnum\minute<10 0\fi\number\minute\the\amorpm}}
\edef\militarytime{\number\hour:\ifnum\minute<10 0\fi\number\minute}

\def\draftlabel#1{{\@bsphack\if@filesw {\let\thepage\relax
      \xdef\@gtempa{\write\@auxout{\string
          \newlabel{#1}{{\@currentlabel}{\thepage}}}}}\@gtempa \if@nobreak
    \ifvmode\nobreak\fi\fi\fi\@esphack} \gdef\@eqnlabel{#1}}
    \def\@eqnlabel{}
\def\@vacuum{}
\def\draftmarginnote#1{\marginpar{\raggedright\scriptsize\tt#1}}

\def\draft{
  \oddsidemargin -.5truein
  \def\@oddfoot{\footnotesize \sl preliminary draft \hfil
    \rm\thepage\hfil\sl\today\quad\militarytime}
  \let\@evenfoot\@oddfoot \overfullrule 3pt
    \let\label=\draftlabel
    \let\marginnote=\draftmarginnote
  \def\@eqnnum{(\theequation)\rlap{\kern\marginparsep\tt\@eqnlabel}%
    \global\let\@eqnlabel\@vacuum}

  }

\makeatletter
\newdimen\normalarrayskip              
\newdimen\minarrayskip                 
\normalarrayskip\baselineskip
\minarrayskip\jot
\newif\ifold             \oldtrue            \def\new{\oldfalse}
\def\arraymode{\ifold\relax\else\displaystyle\fi} 
\def\eqnumphantom{\phantom{(\theequation)}}     
\def\@arrayskip{\ifold\baselineskip\z@\lineskip\z@
     \else
     \baselineskip\minarrayskip\lineskip2\minarrayskip\fi}
\def\@arrayclassz{\ifcase \@lastchclass \@acolampacol \or
\@ampacol \or \or \or \@addamp \or
   \@acolampacol \or \@firstampfalse \@acol \fi
\edef\@preamble{\@preamble
  \ifcase \@chnum
     \hfil$\relax\arraymode\@sharp$\hfil
     \or $\relax\arraymode\@sharp$\hfil
     \or \hfil$\relax\arraymode\@sharp$\fi}}
\def\@array[#1]#2{\setbox\@arstrutbox=\hbox{\vrule
     height\arraystretch \ht\strutbox
     depth\arraystretch \dp\strutbox
     width\z@}\@mkpream{#2}\edef\@preamble{\halign
\noexpand\@halignto
\bgroup \tabskip\z@ \@arstrut \@preamble \tabskip\z@ \cr}%
\let\@startpbox\@@startpbox \let\@endpbox\@@endpbox
  \if #1t\vtop \else \if#1b\vbox \else \vcenter \fi\fi
  \bgroup \let\par\relax
  \let\@sharp##\let\protect\relax
  \@arrayskip\@preamble}
\def\eqnarray{\stepcounter{equation}%
              \let\@currentlabel=\theequation
              \global\@eqnswtrue
              \global\@eqcnt\z@
              \tabskip\@centering
              \let\\=\@eqncr

 \halign to \displaywidth\bgroup
    \eqnumphantom\@eqnsel\hskip\@centering
    $\displaystyle \tabskip\z@ {##}$%
    \global\@eqcnt\@ne \hskip 2\arraycolsep
         $\displaystyle\arraymode{##}$\hfil
    \global\@eqcnt\tw@ \hskip 2\arraycolsep
         $\displaystyle\tabskip\z@{##}$\hfil
         \tabskip\@centering
    &{##}\tabskip\z@\cr}
\newfont{\hr}{msbm10}
\newfont{\ams}{msam10}

\def\be{\begin{eqnarray}}
\def\ee{\end{eqnarray}}

\def\beq{\begin{equation}}
\def\eeq{\end{equation}}
\def\ba{\beq\new\begin{array}{c}}
\def\ea{\end{array}\eeq}
\def\be{\ba}
\def\ee{\ea}

\def\p{\partial}
\def\tr{{\rm tr}\,}
\def\Tr{{\rm Tr}\,}

\def\l[{\phantom.[}
\def\W{\hat W}

\newdimen\linethick  \linethick=0.4pt
\newdimen\hboxitspace    \hboxitspace=5pt
\newdimen\vboxitspace    \vboxitspace=5pt

\def\fr#1{%
\beq\new
\vcenter{
\hrule height\linethick
          \hbox{\vrule width\linethick
                \kern\hboxitspace
                \vbox{\kern\vboxitspace
                      \hbox{$\begin{array}{c}\displaystyle#1
         \end{array}$}%
                      \kern\vboxitspace}%
                \kern\hboxitspace
                \vrule width\linethick}%
          \hrule height\linethick}%
\eeq}

\hoffset= - 1.0in         

\begin{document}

\title{{\bf {On KP-integrable Hurwitz functions} \vspace{.2cm}}
\author{{\bf A. Alexandrov$^{a,b,c,}$}\footnote{alexandrovsash@gmail.com}, {\bf A. Mironov$^{d,c,e,}$}\footnote{mironov@itep.ru; mironov@lpi.ru}, {\bf A. Morozov$^{c,e,}$}\thanks{morozov@itep.ru} and {\bf S. Natanzon$^{f,g,}$}\thanks{natanzons@mail.ru}}
\date{ }
}

\maketitle

\vspace{-6.0cm}

\begin{center}
\hfill FIAN/TD-05/14\\
\hfill ITEP/TH-13/14\\
\end{center}

\vspace{4.2cm}

\begin{center}

$^a$ {\small {\it Freiburg Institute for Advanced Studies (FRIAS), University of Freiburg, Germany }}\\
$^b$ {\small {\it Mathematics Institute, University of Freiburg, Germany }}\\
$^c$ {\small {\it ITEP, Moscow 117218, Russia}}\\
$^d$ {\small {\it Lebedev Physics Institute, Moscow 119991, Russia }}\\
$^e$ {\small {\it National Research Nuclear University MEPhI, Moscow 115409, Russia }}\\
$^f$ {\small {\it National Research University Higher School of Economics,
Moscow, Russia}}\\
$^g$ {\small {\it Laboratory of Quantum Topology, Chelyabinsk State University, Brat'ev Kashirinykh street 129,
Chelyabinsk 454001, Russia}}\\
\end{center}

\vspace{1cm}

\begin{abstract}
There is now a renewed interest \cite{GJ}-\cite{Kaz}
to a Hurwitz $\tau$-function,
counting the isomorphism classes of Belyi pairs,
arising in the study of equilateral triangulations
and  Grothiendicks's {\it dessins d'enfant}.
It is distinguished by belonging to a particular family of
Hurwitz $\tau$-functions,  possessing conventional Toda/KP integrability properties.
We explain how the variety of recent observations about this function
fits into the general theory of matrix model $\tau$-functions.
All such quantities possess a number of different descriptions,
related in a standard way: these include Toda/KP integrability, several kinds
of $W$-representations (we describe four), two kinds of integral (multi-matrix model) descriptions
(of Hermitian and Kontsevich types),
Virasoro constraints, character expansion, embedding into generic
set of Hurwitz $\tau$-functions and relation to knot theory.
When approached in this way,
the family of models in the literature has a natural extension,
and additional integrability with respect to associated new time-variables.
Another member of this extended family is the Itsykson-Zuber integral.
\end{abstract}

\section{Introduction}

Hurwitz $\tau$-function \cite{MMN1,Hurtau}
is a new  important subject of theoretical physics,
which seems relevant to description of non-perturbative phenomena
beyond $2d$ conformal field theory, actually beginning from the
$3d$ Chern-Simons and knot theory, see \cite{Slep}.
In general, Hurwitz $\tau$-functions do not belong \cite{Hurtau} to a narrower
well-studied class of KP/Toda $\tau$-functions, i.e. are not
straightforwardly reducible to free fermions
($\widehat{U(1)}$ Kac-Moody algebras) and Plucker relations
(the Universal Grassmannian).
However, the special cases, when they ${\it do}$, help to
establish links between the known and unknown, and are very
instructive for development of terminology and research directions.
A particular case of the previously known example of this type \cite{OS}
was recently considered  again in \cite{GJ}-\cite{Kaz}
and finally seems to attract reasonable attention.
In the present paper we further extend it and consider
from the perspective of the modern $\tau$-function theory,
thus slightly broadening the consideration in those papers.

In systematic presentation, the story begins from the
celebrated formula \cite{Dij} for the Hurwitz numbers,
\be
{\cal N}_{\Delta_1,\ldots,\Delta_{k}}
= \sum_R d_R^{2}\, \varphi_R(\Delta_1)\ldots\varphi_R(\Delta_{k})
\label{Hurnum}
\ee
which expresses them through the properly normalized
symmetric-group characters $\varphi_R(\Delta)$.
Here $\Delta_1,\ldots,\Delta_{k}$ and $R$ are Young diagrams and $d_R$ is the dimension of representation $R$ of the
symmetric group $S_{|R|}$ divided by $|R|!$, \cite{Mac}.
The ordinary Hurwitz numbers
(counting ramified coverings of the Riemann sphere with
ramifications of a given type)
arise when all $\Delta_1,\ldots,\Delta_{k}$ have the same size
(the same number of boxes), then the sum in (\ref{Hurnum})
goes over $R$ of the same size.
If the size $|\Delta|>|R|$, then $\varphi_R(\Delta)=0$,
if $|\Delta|<|R|$, then
\be
\varphi_R(\Delta) = \frac{(|R|-|\Delta|+k)!}{k!(|R|-|\Delta|)!}\ \varphi_R(\Delta,1^{|R|-|\Delta|})
\ee
where at the r.h.s. $|R|-|\Delta|$ lines of unit length is added to the
Young diagram $\Delta$, and $k$ is the number of lines of unit length in the diagram $\Delta$.
See \cite{MMN1,MMNWDVV} and especially \cite{Hurtau} for more details about all this.

\bigskip

The symmetric group characters $\varphi_R(\Delta)$ are related to the linear group ones
(the Schur functions)
\be
\chi_R[X] = \chi_R\{p\}\Big|_{p_n = \Tr X^n}
\label{schurs}
\ee
as follows \cite{Mac}
\be
\chi_R\{p\} = \sum_\Delta d_R\varphi_R(\Delta)p_\Delta \cdot \delta_{|R|,|\Delta|}
\label{livsyr}
\ee
or \cite{MMNWDVV}
\be
\chi_R\{p_m+\delta_{m,1}\} = \sum_\Delta d_R\varphi_R(\Delta)p_\Delta
\label{livsy}
\ee
The difference between the two is that in (\ref{livsyr}) the sum goes only over $|\Delta|$ of size $|R|$, while in (\ref{livsy})
there is no restriction.
For a Young diagram $\ \Delta: \ = \delta_1\geq\delta_2\geq\ldots\geq \delta_{l(\Delta)}$,
which is an ordered partition of $|\Delta|$ into a sum of $l(\Delta)$ integers $\delta_i$,
associated is the multi-time variable
\be
p_\Delta = p_{\delta_1}p_{\delta_2}\ldots p_{\delta_{l(\Delta)}}
\label{pDelta}
\ee
In the particular case when all $p_n$ are the same, $p_n=N$, i.e. when $X$ is an $N\times N$
unit matrix, $X=I_N$, eq.(\ref{livsyr}) provides $\varphi$-decomposition of the
dimensions $D_R(N)$ of the irreducible representation $R$ of the Lie algebra $gl(N)$
\be
\boxed{
D_R(N) = \chi_R[I_N] = \sum_{\Delta}^{\phantom{5}}  d_R \varphi_R(\Delta) N^{l(\Delta)}
\ \delta_{|R|,|\Delta|}
}
\label{Dviavarphi}
\ee
The standard definition of these dimensions is
the celebrated hook formula \cite{Mac}
\be
{D_R(N)\over d_R}=\prod_{i,j\in R}(N+i-j)=\prod_i{(\lambda_i+N-i)!\over (N-i)!}
\label{Dviapoints}
\ee
In fact, for study of integrability important is just the fact that all $p_n$
are the same, and in
what follows we mostly use the letter $u$ instead of $N$,
to downplay association with the representation dimensions and
emphasize that $u$ does not need to be a positive integer.

\bigskip

Combining (\ref{Hurnum}) and (\ref{livsyr}), it is natural to consider the
generating function \footnote{This definition could depend slightly on whether one imposes restrictions like $|\Delta_i|=|\Delta_j|$
and $|R|=|\Delta_i|$ in the sums.}
\be
h_k\{p^{(1)},\ldots,p^{(k)}\} =
\sum_{\Delta_1,\ldots,\Delta_{k}}
{\cal N}_{\Delta_1,\ldots,\Delta_{k}}\,
p_{\Delta_1}\ldots p_{\Delta_k}
= \sum_R d_R^2 \ \prod_{i=1}^{k} \frac{\chi_R\{p^{(i)}\}}{d_R}
\ee

It is well known that for $k=1$ and $k=2$ these $h$-functions are KP and Toda lattice
$\tau$-functions respectively; moreover, they are trivial $\tau$-functions:
\be
h_1\{p\} = \sum_R d_R\chi_R\{p\} = e^{p_1},  \\
h_2\{\bar p,p\} = \sum_R \chi_R\{\bar p\}\chi_R\{p\} =
\exp \left(\sum_m \frac{1}{m}\bar p_mp_m\right)
\ee
It is also known \cite{Hurtau} that for $k\geq 3$ with generic $p^{(i\geq 3)}$
these $h$-functions do {\it not} belong to the KP/Toda family as functions of
$\{p^{(1)}\}$ or $\{p^{(1)},p^{(2)}\}$.
However, of course, this {\it can} happen for particular choices of $\{p^{(i\geq 3)}\}$,
and they {\it do},
provided all $p^{(i)}_m=u^{(i)}$ for all $m$.

In other words, making use of (\ref{Dviavarphi}) we restrict $h$-functions to
more specific  generating functions:
\be
\boxed{
Z_{(k,n)}(s,u_1,\ldots,u_n\,|\,p^{(i)}) =
\sum_R s^{|R|} d_R^{2-k-n}  \left( \prod_{i=1}^k \chi_R\{p^{(i)}\}\right)
\left( \prod_{i=1}^n D_R(u_i)\right)
}
\label{Zotau}
\ee
at $k=1,2$, which, given their origin and properties, we call hypergeometric (following \cite{OS})
Hurwitz $\tau$-functions.
The formally continued to negative values
$(2,-1)$ member of this family  ${Z}_{(2,-1)}$ is the celebrated Itsykson-Zuber integral:
\be
{Z}_{(2,-1)}\{\bar p, p\} = \sum_R \frac{d_R\chi_R[X]\chi_R[Y]}{D_R(N)} = J_{IZ}(N)
\ee
with $p_n=\tr X^n$ and $\bar p_n=\tr Y^n$ (see eq.(77) in \cite{UnInt}), note that
for representations $R$ with $D_R(N)=0$ the characters in the numerator are also vanishing,
and these $R$ do not contribute to the sum.
For $(1,0)$ and $(1,1)$ we get just the
trivial exponentials
\be
Z_{(1,0)}= \sum_R s^{|R|} d_R\chi_R\{p\} = e^{sp_1}
\ee
\vspace{-0.4cm}
and
\vspace{-0.2cm}
\be
Z_{(1,1)}= \sum_R s^{|R|} D_R(N)\chi_R\{p\} =
\exp\left(N\sum_{m=1}^N{s^mp_m \over m}\right)
\label{Z2}
\ee
The particular case $Z_{(1,2)}$ of generating numbers of isomorphism classes of the
Belyi pairs was studied in \cite{Zog,AC,Kaz}.\footnote{
Belyi pair describes a complex curve as a covering of $CP^1$,
ramified at just three points $0,1,\infty$ (the pair is the curve $C$ and the mapping
$C\longrightarrow CP^1$).
According to G.Belyi and A.Grothendieck \cite{Bel}, existence of such
representation is necessary and sufficient for arithmeticity of the curve and arithmetic curves
are in one-to-one  correspondence with the equilateral triangulations ({\it dessins d'enfant}).
Thus, enumeration of Belyi pairs is a typical matrix model problem (see more on relations between
counting the Belyi maps, Hurwitz numbers and matrix models in \cite{BHM}),
though equivalence of matrix model \cite{Wit,Kon} and sum-over-metrics descriptions \cite{Pol},
proved in \cite{MMMeq,Wit2} on the lines of \cite{2dg,D,FKN} remains a big mystery
from the point of view of the complicated embedding of moduli space of arithmetic curves
into the entire moduli space, see \cite{LevMor} and, for a related consideration, \cite{Gop}.
The Belyi pairs are enumerated by the triple Hurwitz numbers ${\cal N}_{\Delta_0,\Delta_1,\Delta_\infty}$,
but no adequate language is still found to describe the full generating function
$h_3\{p^{(1)},p^{(2)},p^{(3)}) = \sum_R
d_R^{-1}\chi_R(p^{(1)})\chi_R(p^{(2)})\chi_R(p^{(3)})$, see \cite{Hurtau}.
The suggestion of \cite{Zog} was to sacrifice any details about $\Delta_0$ and $\Delta_1$
and keep only information about the numbers $l(\Delta_0)$ and $l(\Delta_1)$
of unglued sheets of the covering over $0$ and $1$: then such special
generating function $Z_{(1,2)}$ is obviously a KP $\tau$-function.
In fact, it is enough to do so just at one (not obligatory two) of the three points:
$Z_{(2,1)}$ is also a conventional Toda lattice $\tau$-function.
Presentation of standard results about these quantities and their multi-point
counterparts is the purpose of the present paper.
As to triple coverings, enumeration is the simplest, but not the most interesting
part of the story. An explicit construction of the Belyi functions is extremely hard:
for relatively vast set of examples see \cite{Sha}.
A crucial problem of string theory remains expressing the Mumford measure and its
constituents (determinants of $\bar\p$ operators) for arithmetic curves
through combinatorial triple $\Delta_1,\Delta_2,\Delta_3$.
}

In fact, {\bf models $Z_{(1,n)}$ with $n>2$ are far more interesting}.
This becomes obvious already for $N=1$, when
only symmetric diagrams $R=[m]$ contribute, with $D_{[m]}(N=1)=1$ and
$d_{[m]} = 1/m!$, so that (\ref{Zotau}) turns into a simple series
\be
Z_{(1,n)}\Big({\rm all}\ u_i=1\Big) =
\sum_{m=0}^\infty s^m \chi_{[m]}\{p\} d_{[m]}^{1-n}D_{[k]}^n =
\sum_{m=0}^\infty (m!)^{n-1} s^m\chi_{[m]}\{p\} =
\sum_{m=0}^\infty \underline{(m!)^{n-2} (sp_1)^m} + O(p_2,\ldots)
\label{hierser}
\ee
The underlined series is nicely convergent for $n=1$ and $n=2$,
while  for $n>2$ it is asymptotic series, defined up to
non-perturbative corrections.
For $n=3$ we get the archetypical example:
\be
\sum_m   m!\cdot s^m
\vspace{-0.2cm}
\ee
where non-perturbative ambiguity is proportional to
\vspace{+0.1cm}
\be
\oint \frac{e^{-x}dx}{1-xs} = \frac{e^{-1/s}}{s}
\ee
This example appears in the study of $Z_{(1,3)}$.
The usual way to handle the series like (\ref{hierser})
is the integral transformation:
\be
f(s) = \sum_m a_m s^m \ \longrightarrow \ F(s) =\sum_m  a_mm!\cdot s^m
=\frac{1}{s} \int_{x_+} e^{-x/s} f(x) dx
\ee
For generic $N$ this formalism turns into the theory of Kontsevich-like
models.

\bigskip

Of course, (\ref{Zotau}) are very special, besides they are $\tau$-functions \cite{OS,GJ},
they actually belong to the class of
{\it matrix model $\tau$-functions} \cite{UFN23}.
This not-yet-rigourously-defined class is characterized by coexistence
of a wide variety of very different representations and properties
\cite{MorMaMo}:

-- they are KP/Toda $\tau$-functions,

-- they possess integral (``matrix-model") representations of
``ordinary" and Kontsevich types,

-- they satisfy Virasoro- or W-like constraints (possess a
$D$-module representation and obey the AMM/EO topological recursion \cite{AMMEO}),

-- they possess various $W$-representations \cite{Wreps}, including ones via Casimir operators
and via cut-and-join operators,

-- they possess special linear decompositions into linear- and symmetric-group
characters,

-- they are Hurwitz $\tau$-functions.

\noindent
The purpose of this paper is to describe all
these properties within the context of
the hypergeometric Hurwitz $\tau$-functions (\ref{Zotau}).

\bigskip

For illustrative purposes and to avoid notational confusions
we list the simplest examples of
dimensions (\ref{Dviapoints}), linear group characters $\chi_R\{p\}$, and
appropriately normalized symmetric group
characters $\varphi_R(\Delta)$ from \cite{MMN1}:
$$
\begin{array}{c|c||c|c|| c|cc|ccc|c}
R   & D_R(N)/d_R & \chi_R\{p\}&d_R & \varphi_R(1) & \varphi_R(2) & \varphi_R(11) & \varphi_R(3)
& \varphi_R(21) & \varphi_R(111)&\ldots \\
\hline
\l[1] &   N & p_1 & 1&1 &&&&& \\
\l[2] & N(N+1) & \frac{p_2+p_1^2}{2} & \frac{1}{2} & 2 & 1 & 1 &&&&  \\
\l[11] & N(N-1) & \frac{-p_2+p_1^2}{2} & \frac{1}{2} & 2 & -1 & 1 &&&& \\
\l[3] & N(N+1)(N+2) & \frac{2p_3+3p_2p_1+p_1^3}{6} &\frac{1}{6} & 3 &3&3&2&3&1& \\
\l[21] & (N-1)N(N+1) & \frac{-p_3+p_1^3}{3} & \frac{1}{3} & 3 & 0&3&-1&0&1& \\
\l[111] & N(N-1)(N-2) & \frac{2p_3-3p_2p_1+p_1^3}{6} & \frac{1}{6} & 3 & -3&3&2&-3&1&\\
\ldots
\end{array}
\label{charlist}
$$

\bigskip

\section{Representation via  cut-and-join operators}

The linear group characters (Schur functions) $\chi_R\{p\}$
are common eigenfunctions of the set of commuting
generalized cut-and-join operators \cite{MMN1},
and symmetric group characters $\varphi_R(\Delta)$
are their corresponding eigenvalues:
\be
\hat W_\Delta \chi_R = \varphi_R(\Delta)\chi_R
\label{Wef}
\ee
What we need in (\ref{Zotau}) are rather operators with slightly different
eigenvalues:
\be
\hat O(u) \chi_R = \frac{D_R(u)}{d_R} \chi_R
\ee
However, eq.(\ref{Dviavarphi}) allows one to make them easily from $\hat W_\Delta$:
\be
\boxed{
\hat O(u) = \sum_\Delta^{\phantom{5}} u^{l(\Delta)} \hat W_\Delta \hat P_{|\Delta|}
}
\label{hatO}
\ee
where $\hat P_{|\Delta|}$ is a projector, selecting the Young diagrams of the size $|\Delta|$,
\be
\hat P_{|\Delta|} = \oint \frac{dz}{z}\ z^{- |\Delta|+\hat L_0}
\ee
with
\be
\hat L_0 = \sum_n np_n\frac{\p}{\p p_n}
\ee
so that
\be
\hat P_{|\Delta|}\chi_R = \chi_R \, \delta_{|R|,|\Delta|}
\ee
and $\hat W_\Delta$ are the general cut-and-join operators from \cite{MMN1}.

\bigskip

Thus
\vspace{-0.5cm}
\be
Z_{(1,n)}(s,u_1,\ldots,u_n|p) = \left(\prod_{i=1}^k\hat O(u_i)\right) e^{sp_1},  \\
{Z}_{(2,n)}(s,u_1,\ldots,u_n|\bar p,p) = \left(\prod_{i=1}^n\hat O(u_i)\right)
\exp\left(\sum_m \frac{s^m}{m}p_m\bar p_m\right)
\label{WrepO}
\ee
These are actually the $W$-representations \cite{Wreps} of the
$\tau$-functions (\ref{Zotau}), because $\hat O(u)$ are, in fact, elements of the
integrability-preserving $GL(\infty)$ group.
However, this is not quite so obvious: operator (\ref{hatO}) does not have
a form where this property is obvious.
In fact, one can make a triangular transformation in (\ref{hatO}) and get rid
of projector operators $\hat P_{|\Delta|}$:
$$
\hat O(u) = u^{\hat L_0}\left(1 + \frac{\hat W_2}{u} + \frac{\hat W_3 + \hat W_{22}}{u^2}
+ \frac{\hat W_4+\hat W_{32}+\hat W_{222}}{u^3} + \frac{\hat W_5+\hat W_{42}+\hat W_{33}+\hat W_{322}
+ \hat W_{2222}}{u^4} \ldots \right) =
$$
or
\be
\boxed{\hat O(u)= u^{\hat W_1}\ {\sum_{\Delta}}^\prime u^{l(\Delta)-|\Delta|} \,\hat W_\Delta}
\label{OvsW}
\ee
where sum goes over all diagrams containing no lines of unit length (we denote this restriction by prime).

Since, say \cite{MMN1},
$\hat W_{22} = \frac{1}{2}\Big(\hat W_2^2 - 3\hat W_3 - \hat W_{11}\Big)$,
this expressions has chances to be exponentiated. In this case, the exponent should contain
even less types of operators,
to provide an element from $GL(\infty)$:
it should actually be \cite{Hurtau} a {\it linear} combination of Casimir operators.
We shall now demonstrate this.

\section{Representation via Casimir operators}

We want to find an exponential representation of the operator $\hat O(u)$,
and what we know is that the eigenvalues of $\log \hat O(u)$
are logarithms of (\ref{Dviapoints}).
More precisely, we need the $1/N$-expansion of
\be
\log \left(\frac{D_R(N)}{N^{|R|}\cdot d_R}\right) =
\sum_{(i,j)\in R} \log \left(1+\frac{i-j}{N}\right)
= \sum_{m=1}^\infty \frac{(-)^{m+1}}{N^m \cdot m}\, \tilde\sigma_R(m+1)
\ee
\vspace{-0.2cm}
where
\vspace{-0.2cm}
\be
\tilde\sigma_R(m+1) = \sum_{(i,j)\in R} (i-j)^m=\sum_{k=0}^m{m!\over k!(m-k)!}
\sum_{j=1}^{l(\Delta)}\left((-j)^{m-k}\sum_{i=1}^{r_j}i^k\right)
\ee
In fact, one can easily check that these quantities are
linear combinations of the eigenvalues $\sigma(m)$ of the
Casimir operators \cite{Zhe},
\be
\hat C_m \chi_R = \sigma_R(m) \chi_R
\ee
which are given by
\be
\sigma_R(m) = \frac{1}{m}\sum_{j=1}^{l(R)} \Big((r_j -j+1/2)^m - (-j+1/2)^m\Big)
\label{staCas}
\ee
In particular,
\be
\sigma_R(1) = \sum_i r_j = \sum_{(i,j)\in R} 1 = \tilde \sigma_R(1),  \\
\sigma_R(2) = \frac{1}{2}\sum_{j=1}^{l(R)} r_j(r_j-2j+1)  =
\sum_{j=1}^{l(R)}\left(\frac{r_j(r_j+1)}{2} - jr_j\right)
= \sum_{(i,j)\in R} (i-j) = \tilde\sigma_R(2), \\
\ldots
\ee
However, for higher $m$ relations are a little more involved:
\be
\tilde\sigma_R(m) = \sigma_R(m) - \sum_{k=1} \frac{(m-1)!}{(2k)!(m-1-2k)!}\Big(1 - 2^{1-2k}\Big)
B_{2k}\cdot\sigma_R(m-2k)
\label{sigmatilde}
\ee
The sum has finite number of items, $k<\frac{m}{2}$, and $B_{2k}$ are the Bernoulli numbers,
\be
\sum_n \frac{B_mt^m}{m!} = \frac{te^t}{e^t-1}, \ \ \ \ \ {\rm or} \ \ \ \ \ \  \
\sum_n {B_{2m}t^{2m}\over (2m)!}={te^t\over e^t- 1}-1-\frac{t}{2}
\ee
$B_1=\frac{1}{2},\ B_2=\frac{1}{6},\ B_4=-\frac{1}{30},\ B_6=\frac{1}{42},\  B_8=-\frac{1}{30},\
B_{10} = \frac{5}{66},\ B_{12} = -\frac{691}{2730}, \ B_{14}=\frac{7}{6},
\ B_{16} = -\frac{3617}{510}, \ \ldots$

\bigskip

What is important about the Casimir operators is that they contain single
sums over $j$, and this property guarantees integrability \cite{Hurtau}.
It is of course preserved by {\it linear} combinations,
i.e. $\hat{\widetilde C}_n$ with the eigenvalues $\tilde\sigma(n)$
are as good from this point of view as $\hat C_n$ with the eigenvalues $\sigma(n)$.

\bigskip

Thus we obtained the desired exponential representation of the operators
\be
\boxed{
\ \ \
\hat O(u) = u^{\hat L_0} \exp \left\{  \sum_{m=1}^\infty
\frac{(-)^{m+1}}{u^m \cdot m}\, \hat{\widetilde C}_{m+1} \right\}
\ \ \
}
\label{OCas}
\ee
Moreover, when there are many $u$ variables, one can simply consider
them as the Miwa-like reparametrisation of a new type of variables,
\be
\eta_m = \frac{(-)^{m+1}}{m}\sum_{i=1}^n u_i^{-m}, \ \ \ \ \ \ \ \ \
\eta_0 = \sum_{i=1}^n \log u_i
\ee
and the function (\ref{WrepO}) becomes also a function of these
additional time-variables $\eta$:
\be
Z_{(1,n)}(s,u_1,\ldots,u_n|p) = \left(\prod_{i=1}^n\hat O(u_i)\right) e^{sp_1} =
\exp\left(\sum_{m=0}^\infty \eta_m\, \hat{\widetilde C}_{m+1}\right)\cdot e^{sp_1}
\label{Wrep01}
\ee
This function, as a function of the variables $\eta_k$, is very similar to the $\tau$-function \cite{Hurtau}:
\be
Z=\exp\left(\sum_{m=0}^\infty \bar\eta_m\, \hat{C}_{m+1}\right)\cdot e^{sp_1}
\ee
where the variables
\be
\bar\eta_m=\sum_{i=1}^n u_i^{-m},
\ee
related to $\sigma_R(m)$, are the linearly transformed variables $\eta_m$. In spite of this similarity, two functions are not connected with each other by a relation, describing equivalent integrable hierarchies \cite{eqh}. In particular, change of the basis (\ref{sigmatilde}), which relates the operators $\hat{C}_k$ with $\hat{\tilde C}_k$, is not given by a change of the spectral parameter, see e.g. \cite{Alex} for more details.

Explicit relation between (\ref{OCas}) and (\ref{OvsW}) is an interesting
exercise, concerning commutative algebra of cut-and-join operators
and their relation to the Casimir operators.
It can be easily checked in the lowest orders of the $u^{-1}$-expansion
with the help of multiplication table from \cite{MMN1}.

\section{$Z_{(2,n)}$ as a $\tau$-function of Toda lattice}

Eq.(\ref{OCas}) immediately implies that $Z_{(2,n)}$ is a Toda lattice $\tau$-function
(thus, $Z_{(1,n)}$ is a KP $\tau$-function). Indeed,
according to  \cite{Hurtau} the exponential of
{\it linear} combinations of the Casimir operators belongs to $GL(\infty)$
which preserves the KP/Toda integrability. In case of $Z_{(2,n)}$
the product of the $GL(\infty)$ operators (\ref{OCas}) acts on the trivial $\tau$-function
$\ \exp\left(\sum_m \frac{s^mp_m\bar p_m}{m}\right)$.

Still, there are many other ways to demonstrate that $Z_{(2,n)}$
is a $\tau$-function of the Toda lattice hierarchy.
The most important is the free-fermion approach of \cite{DJKM}
and closely related determinant formulas, see \cite{versus,UFN23}.
From the point of view of Hurwitz theory, the basic well-known fact is that
the character expansion
\be
\sum_R g_R\,\chi_R(p)
\ee
is a KP $\tau$-function {\it iff} coefficients $g_R$ satisfy the Pl\"ucker relations,
of which the generic solution is
\be
g_R = \det_{ij}\Big(F(r_i-i,j)\Big)
\label{KPint}
\ee
with arbitrary function $F$ of two variables.

Likewise, according to \cite{Taka}
\be\label{OS}
\tau_n(p,\bar p|f)=\sum_{R,R'} f_{R,R'}(n)\,\chi_R(p)\chi_{R'}(\bar p)
\ee
is a Toda lattice $\tau$-function, {\it iff}
\be
f_{R,R'}(n) = \det_{ij\le n}\Big(F(r_i-i,r'_j-j)\Big)
\label{Todaint}
\ee
Parameter $n$ here plays a role of the Toda zero-time $p_0$.

A particular class of solutions of this type
is provided by a much simpler diagonal coefficients $f_{R,R'}(n)$ \cite{OS,GJ}
\be
f_{R,R'}(u)=\delta_{R,R'}\prod_{i,j\in R} f(u+i-j)
\label{OSint}
\ee
where $f(x)$ is an arbitrary function of a single variable.
This class of $\tau$-functions of the Toda lattice hierarchy
explicitly given by the free-fermion average
\be
\tau_n(p,\bar p|f)=\left<n\left|e^{H(p)}e^{\sum T_m :\psi^*_m\psi_m:}e^{\bar H(\bar p)}\right|n\right>
\ee
where the normal ordering is defined w.r.t. the zero vacuum:
$:\psi^*_m\psi_m:=\psi^*_m\psi_m-<0|\psi^*_m\psi_m|0>$ and the coefficients
$T_k$ are introduced via $f(k)=e^{T_{k-1}-T_k}$ with $T_{-1}=0$.
More explanations of the notation see in \cite{DJKM,versus,OS}. This $\tau$-function
was named   {\it hypergeometric}  in \cite{OS}. In particular, from (\ref{sigmatilde}) is follows that the operators $\hat O(u)$ yield the coefficients precisely of the this form, thus the functions $Z_{(2,n)}$ belong to this class.

In fact, one can even restrict the sum in (\ref{OS}) to the diagrams with no more than $n$ lines,
where $n$ is the zero-time:
\be
\tilde\tau_n(p,\bar p|f)=\sum_{R:\ l(R)\leq n} f_R(n)\chi_R(p)\chi_R(\bar p)
\ee
it is still a Toda lattice $\tau$-function \cite{Hurtau}.

\bigskip

The generic Hurwitz $\tau$-function
\be
h(p^{(1)},\ldots,p^{(k)}|\beta) =
\exp\left(\sum_\Delta \beta_\Delta \hat W_\Delta\right)
\sum_R d_R^{2-k}\chi_R\{p^{(1)}\}\ldots \chi_R\{p^{(k)}\}
\ee
does not satisfy criteria (\ref{KPint}) and (\ref{Todaint}) as a function of any
time or time pairs, see \cite{Hurtau} for a detailed consideration
(it is not even clear if it fits into the wide class of the non-Abelian $\tau$-functions
of \cite{KLMMM}).
Notable exceptions are the cases when $k=1,2$ and when
$\beta_\Delta$ are adjusted to provide any {\it linear} combination of
the standard Casimir operators (\ref{staCas}), which are
nicknamed as {\it complete cycles} in \cite{Ok}.
The functions (\ref{Zotau}) use additional freedom (\ref{OSint}) to enlarge $k$,
but keeping $p^{(3)},\ldots,p^{(k)}$ very special: constant.
This corresponds to choosing $f(x)=\prod_{i=1}^k (x+u_i)$ in (\ref{OSint})
while the $s$-dependence is introduced by the rescaling $p_k\to s^kp_k$.

Of course, this $Z_{(2,n)}(u_1,\ldots,u_n\,|\,p,\bar p)$ is a very special kind of a
 lattice $\tau$-function.
In particular, it possesses a simple integral representation in the form of
eigenvalue matrix model (as foreseen already in \cite{OS}).
We construct such representations in the generic case in the next section,
and then consider particular more explicit examples.

\section{Matrix model representations}

Making use of orthogonality condition \cite[eq.(3.1)]{CJR},\footnote{
The simplest way to prove (\ref{2mort}) is to make use of formula from Fourier theory
\be\label{bas}
\left.\int dxdyf(x)g(y)e^{-xy}=f\left({\partial\over\partial x}\right)g(x)\right|_{x=0}
\ee
where the $x$-integral goes over the real axis, and the $y$-integral runs over the imaginary one.
Now after performing the integration over angular variables and using the Itzykson-Zuber
formula, one obtains the multiple eigenvalue integral
\be
\int dXdY\chi_R(X)\chi_Q(Y) e^{-\tr XY}\sim\int \prod_i \det_{ij} x_i^{N+R_j-j}
\det_{ij} y_i^{N+Q_j-j} e^{-\sum_i x_iy_i}
\ee
where we used the Weyl formula for the characters of linear groups
\be
\chi_R={\det_{ij} x_i^{N+R_j-j} \over\Delta(x)}
\ee
and $\Delta(x)$ is the Van-der-Monde determinant.
Using now formula (\ref{bas}) and
\be
\int \det_{ij} f_i(x_j)\det_{ij} g_i(y_j) \prod_i K(x_i,y_i)=\det_{ij} \int f_i(x)g_j(y)K(x,y)
\ee
one immediately obtains (\ref{2mort}).

This formula can be also described in the pure combinatorics terms using the Feynman diagrams.
The role of propagator here is played by
$\ \left<X_{ij}\, Y_{kl}\right>\ = \delta_{il}\delta_{jk}\,$.
Therefore, the formula reduces to trivial combinatorics:  connecting the free
ends of multi-linear combinations of trace operators.
For example,
\be
 \\
\left< \Tr X \ \Tr Y\right> \ =\delta^{ij}\delta^{kl}\delta_{il}\delta_{jk} = N, \\  \\
\left< \Tr X^2 \ \Tr Y^2\right> \ = 2N^2,    \\
\left< \Tr X^2 \ \Big(\Tr Y\Big)^2\right> \ = 2N,   \\
\left< \Big(\Tr X\Big)^2\ \Big(\Tr Y\Big)^2\right> \ = 2N^2, \  \\  \\
\ldots \ \ \ \
\ee
\begin{picture}(100,0)(-400,-70)
\qbezier(50,2)(19,8)(27,34) \put(48,2.5){\vector(4,-1){2}}
\qbezier(50,-2)(19,-8)(27,-34) \put(26,-28){\vector(1,-4){2}}
\qbezier(-23,36)(0,17)(23,36) \put(21,34){\vector(1,1){2}}
\qbezier(-23,-36)(0,-17)(23,-36) \put(-21,-34){\vector(-1,-1){2}}
\qbezier(-50,2)(-19,8)(-27,34)  \put(-26,28){\vector(-1,4){2}}
\qbezier(-50,-2)(-19,-8)(-27,-34) \put(-48,-2.5){\vector(-4,1){2}}
\put(-8,-2){\mbox{$\Tr X^6$}}
\end{picture}
}
\be
\int\int_{N\times N}  \chi_R[X]\chi_Q[Y]\, e^{i\,\Tr XY} dXdY = \frac{D_R(N)}{d_R}\,\delta_{R,Q}
\label{2mort}
\ee
one can easily rewrite (\ref{Zotau}) in the form of multi-matrix models.
Indeed, from (\ref{2mort}) it follows that
$$
{Z}_{(2,1)}(N|p,\bar p) = \sum_R \frac{D_R(N)}{d_R}\chi_R\{p\}\chi_R\{\bar p\} =
\int\int_{N\times N}  \left(\sum_R\chi_R[X] \chi_R\{p\}\right)
\left(\sum_Q \chi_Q[Y] \chi_Q\{\bar p\}\right)  e^{i\,\Tr XY} dXdY =
$$
\vspace{-0.4cm}
\be
= \int\int_{N\times N} e^{\sum_n \frac{1}{n}p_n\Tr X^n + \sum_n \frac{1}{n}\bar p_n \Tr Y^n}
e^{i\,\Tr XY} dXdY
\label{MaMo3}
\ee
what is just the conventional 2-matrix model, as was already noted in \cite{OS}.

Here we used the relation
\be
\sum_Q \chi_Q[Y]\chi_Q\{\bar p\} = e^{\sum_n  \frac{1}{n}\bar p_n \Tr Y^n}
\ee
which we also need below in the form
\be
\sum_S \chi_S[Y_1]\chi_S[X_2] = e^{\sum_n  \frac{1}{n}\Tr Y_1^n \Tr X_2^n}
= {\rm Det} \Big(I\otimes I -Y_1\otimes X_2\Big)^{-1}
\ee

Similarly to (\ref{MaMo3}),
$$
{Z}_{(2,2)}(N_1,N_2|p,\bar p) = \sum_R \frac{D_R(N_1)D_R(N_2)}{d_R^2}\chi_R\{p\}\chi_R\{\bar p\}=
$$

$$
\!\!\!\!\!\!\!\!\!\!\!\!
=  \sum_S \int\int_{N_1\times N_1}  \left(\sum_R \chi_R\{p\}\chi_R[X_1] \right)
  \chi_S [Y_1]   \, e^{i\,\Tr X_1Y_1}\, dX_1dY_1
   \int\int_{N_2\times N_2} \chi_S [X_2] \left(\sum_Q \chi_Q[Y_2]\chi_Q\{p\} \right)
    \, e^{i\,\Tr X_2Y_2}\, dX_2dY_2 =
$$

$$
= \int\int_{N_1\times N_1} e^{\sum_n \frac{1}{n}p_n\Tr X_1^n}\, e^{i\,\Tr X_1Y_1}\, dX_1dY_1
\int \int_{N_2\times N_2} e^{\sum_n \frac{1}{n}\bar p_n\Tr Y_2^n}\, e^{i\,\Tr X_2Y_2}\, dX_2dY_2
\ \frac{1}{{\rm Det}\Big(I_{N_1}\otimes I_{N_2} - Y_1\otimes X_2\Big)} =
$$
\vspace{-0.4cm}
\be
= \int_{N_1\times N_1}\int_{N_2\times N_2}
d{\cal K}_{N_1}(Y_1|p)\,\frac{1}{{\rm Det}\Big(I_{N_1}\otimes I_{N_2} - Y_1\otimes X_2\Big)}\,
d{\cal K}_{N_2}(X_2|\bar p)
\ee
where generalized Kontsevich measure is defined as \cite{GKM}
\be
d{\cal K}_N(Y|p) = dY \int_{N\times N} e^{\sum_n \frac{1}{n}p_n\Tr X^n}\, e^{i\,\Tr XY}\, dX
\ee

Further,
\vspace{-0.3cm}
\be
{Z}_{(2,3)}(N_1,N_2,N_3|p,\bar p) =
\ \  \ \ \ \ \ \ \ \ \ \ \ \ \ \ \ \ \ \ \ \ \ \ \ \ \ \ \
\ \ \ \ \ \ \ \ \ \ \ \ \ \ \ \ \
 \\  \\
= \int
d{\cal K}_{N_1}(Y_1|p)\,\frac{1}{{\rm Det}\Big(I_{N_1}\otimes I_{N_2} - Y_1\otimes X_2\Big)}
\,e^{i\,\Tr X_2Y_2}\, dX_2dY_2\,
\frac{1}{{\rm Det}\Big(I_{N_2}\otimes I_{N_3} - Y_2\otimes X_3\Big)}\,
d{\cal K}_{N_3}(X_3|\bar p)
\label{Z23Mamo}
\ee
and for generic $k$ we have:
\be
\boxed{{Z}_{(2,n)}\big(N_1,\ldots, N_{n}\big|p,\bar p\big) =
\int
d{\cal K}_{N_1}(Y_1|p)\,
\frac{\prod_{i=2}^{n-1}\, e^{i\,\Tr X_iY_i}\, dX_idY_i}
{\prod_{i=1}^{n-1}
{\rm Det}\Big(I_{N_i}\otimes I_{N_{i+1}} - Y_{i}\otimes X_{i+1}\Big)}\,
d{\cal K}_{N_{n}}(X_{n}|\bar p)}
\ee
One can observe amusing parallels with the conformal matrix models
\cite{confMaMo}, which already have a number of other interesting applications
\cite{confdop}.

One can make the Miwa transformation of time variables $p_m=\Tr \Lambda^{-m}$
in order to transform these matrix integrals to an equivalent form
depending on the external matrix $\Lambda$. Sometimes it turns out very convenient
as we shall see below.

\section{Miwa transformation to Kontsevich matrix models
\label{MiKo}}

Now we make the Miwa transformation of one set of the time variables in $Z_{(2,k)}$
in order to obtain matrix integrals of the Kontsevich type.
This kind of integrals are sometimes more convenient.
In particular, the Virasoro constraints for  $Z_{(1,2)}$ are evident in this representation.

For the sake of simplicity, we consider only $Z_{(2,1)}$ case, a generic case is treated in full analogy.
Thus, we make the Miwa transformation of times $\bar p_m=\Tr\Lambda^{-m}$ in the formula (\ref{MaMo3}),
so that
\be
\exp\left(\sum_n\frac{\bar p_m}{m}\,\Tr Y^m\right)=
{\big(\det \Lambda\big)^{N}\over\det(\Lambda\otimes I-I\otimes Y)}
\ee
Then, the integral becomes
\be
\!
Z_{(2,1)}(N|p,\bar p)
=\!\int dXdY e^{i\,\Tr XY} e^{\sum_m {p_m\over m}\Tr X^m} e^{\sum_m {\bar p_m\over m} \Tr Y^m}\!\!\!
=\!\int dXdY e^{i\,\Tr XY} e^{\sum_m {p_m\over m}\Tr X^m}\!\!
{\big(\det \Lambda\big)^{N}\over\det(\Lambda\otimes I-I\otimes Y)}
\ee
The integral over matrix $Y$ can be easily calculated (to this end, one has first to perform
integration over the angular variables and then make Fourier transform
w.r.t. to the eigenvalues of $Y$), the result reads
\be\label{KI}
Z_{(2,1)}(N|p,\bar p)=\big(-\det \Lambda\big)^{N}\int_{X_+} dX_{N\times N}
e^{\,-\Tr X\Lambda+\sum_m {p_m\over m}\Tr X^m}
\ee
where integral runs over $N\times N$ positive-definite matrices, that is matrices with {\bf positive eigenvalues}.
This follows from the standard Fourier transform:
\be\label{bi}
\int {e^{ixy}\over y-i0}\,dy =2\pi i \theta (x)
\ee

Integral (\ref{KI}) is not yet quite of Kontsevich type: it {\it essentially} depends on
the matrix size $N$ and one can not reach
an arbitrary point in the space of time variables. In order to lift this restriction,
one can add the logarithmic term
which makes the parameter $u$ and the number of integrations independent variables:
\be\label{z21mm}
\boxed{
Z_{(2,1)}(u|p,\bar p)=\big(-\det \Lambda\big)^{u}\int_{X_+} dX_{N\times N}
e^{\,-\Tr X\Lambda+(u-N)\Tr \log X+\sum_m {p_m\over m}\Tr X^m}}
\ee
One can easily check for concrete $N$ that expansion of this integral into $p_k$-series
coincides with $Z_{(2,1)}(u|p,\bar p)$ from (\ref{Zotau}).
Note also that this integral, if considered as a function of time variables
$\bar p_m = \Tr \Lambda^{-m}$,
does not depend\footnote{This integral is independent of $N$ in the following sense.
Calculate the coefficient in front of, say, $p_1p_2$ at different values of $N$:
$$
N=1:\ \ \ \ {u(u+1)(u+2)\over 2\lambda^3}
$$
$$
N=2:\ \ \ \ {u(u+1)(u+2)\over 2}\Big({1\over\lambda_1}+{1\over\lambda_2}\Big)
\Big({1\over\lambda_1^2}+{1\over\lambda_2^2}\Big)
$$
$$
N=3:\ \ \ \ u^2\Big({1\over\lambda_1^3}+{1\over\lambda_2^3}+
{1\over\lambda_3^3}\Big)+{u(u^2+2)\over 2}
\Big({1\over\lambda_1}+{1\over\lambda_2}+{1\over\lambda_3}\Big)
\Big({1\over\lambda_1^2}+{1\over\lambda_2^2}+
{1\over\lambda_3^2}\Big)+{u^2\over 2}\Big({1\over\lambda_1}+{1\over\lambda_2}
+{1\over\lambda_3}\Big)^3
$$
$$
\ldots
$$
All these expressions look different and depending on $N$, but in fact are all equal to
the independent on $N$ polynomial
$$
u^2\bar p_3+{u(u^2+2)\over 2}\bar p_1\bar p_2+{u^2\over 2}\bar p_1^3
$$
}
on $N$, which is the necessary property of Kontsevich integrals \cite{GKM}.

Integral (\ref{z21mm}) was obtained in \cite{AC} within a different approach.
From this formula one immediately obtains a one-matrix model describing
$Z_{(1,2)}(u,v|p)$ at integer points $v=N$.
This can be done in different ways.

One possibility is to put $p_m = v$, then we obtain the double-logarithm
model of \cite{AC}:
\be
\big(-\det \Lambda\big)^{u}\int_{X_+} dX_{N\times N}
e^{\,-\Tr X\Lambda+(u-N)\Tr \log X-v\log(1-X)}
\label{doulog}
\ee
This kind of models were thoroughly investigated in \cite{CP},
still in a moment we will see that (\ref{doulog}) is equivalent to an even better
studied theory.

Another possibility just to put $\Lambda=1$.
Then the result is
\be\label{z12mm}
\boxed{
Z_{(1,2)}(u,v|p)\Big|_{v=N}=\int_{X_+} dX_{N\times N}
e^{\,-\Tr X+(u-N)\Tr \log X+\sum_n {p_n\over n}\Tr X^n}}
\ee
Since this integral goes over only the positive $X_+$, it is equivalent to the model of complex matrices
where $X$ is an obviously positive-definite matrix product $HH^\dag$ \cite{Cmamo,MMMM}:
\be
Z^{\bf C}=\int dHdH^\dag e^{\Tr V(HH^\dag)}\sim\int\prod_i dh_i^2\Delta^2(h_i^2)e^{\sum_i V(h_i^2)}
\label{coma}
\ee
where $V(X)$ is arbitrary potential of the matrix model, $h_i^2$ are eigenvalues of
$HH^\dag$ and $\Delta(h)$ is the Van-der-Monde determinant.
{\bf Thus, $Z_{(1,2)}(u,v|p)$  from \cite{Zog,AC,Kaz}, and, hence, the double-logarithm model
(\ref{doulog}) is nothing but the well-known complex matrix model}.
Among other things, this immediately implies the Virasoro constraints for $Z_{(1,2)}(u,v|p)$.

In a similar way one can make the Miwa transformation of one set of times and perform integration like (\ref{bi}) in order
to obtain from (\ref{Z23Mamo}) a {\it two}-matrix model representation of
$Z_{1,3}$:
\be\label{Z13}
Z_{1,3}(u,v,w|p)\Big|_{u=N} \sim
\int_{X_+}dX_{N\times N} \int_{Y_+}dY_{N\times N} \exp\Big(-\Tr Y^{-1} - i\, \Tr XY
+ \sum_m \frac{p_m}{m}\Tr X^m  +\\
+(v-N)\Tr\log X + (v-w-N)\Tr\log Y\Big)
\ee
where we assume that $N\le v\le w$ (hence the asymmetry of the integral w.r.t. interchanging
$v$ and $w$), otherwise the integrals diverge.
From experience in \cite{Virmult} and \cite{GKMU} it comes with no surprise that this $Z_{(1,3)}$
satisfies $\widetilde W^{(3)}$ constraints. In these constraints the values of $u$, $v$ and $w$ are arbitrary, and
the symmetry is restored (see (\ref{Wc})).

\section{Virasoro/$\widetilde W$ constraints}

The simplest way to obtain Virasoro/W constraints for $Z_{(k,n)}$ is to construct
the loop equations (Ward identities) of the corresponding matrix models, which are associated with arbitrary changes of integration variables in the matrix integral.
The Ward identities for the two-matrix model describing $Z_{(2,n)}$ are quite involved and
are expressed in terms of the $\widetilde W_{\infty}$-algebra of ref.\cite{Virmult}.
However, when one set of times is eliminated things simplify a lot.
In particular, when only $l$ first $\bar p_{i}$, $i\le l$, are non-vanishing, the constraints imposed on
$p$-dependence involve only $\widetilde W^{(i)}$-operators with $i\leq l$ \cite{Virmult}.
As we now see, the same seems  true for $Z_{(1,n)}$ models, where all $\bar p$ are non-vanishing,
but the same.
This result can imply additional kinds of matrix-model representations for $Z_{(1,n)}$.

To begin with, $Z_{(1,1)}(u|p) = \exp\left(\sum_{m=1}^\infty \frac{us^mp_n}{m}\right)$ satisfies
\be
\boxed{\
\left(\hat J_m^{\bf C}- \frac{m+1}{s}\frac{\p}{\p p_{m+1}}\right)
\,Z_{(1,1)}(u\,|\,p) =
0,
\ \ \ \ m\geq 0
\ }
\label{lincon}
\ee
with
\be
\hat J_m^{\bf C} = m\frac{\p}{\p p_m}
\ee

The next model $Z_{(1,2)}$ is equivalent to
the complex one-matrix model (\ref{z12mm}), for which
the Ward identities are just the Virasoro constraints, derived in \cite{MMMM}:
\be
\boxed{\
\left(\hat L_m^{\bf C}- \frac{m+1}{s}\frac{\p}{\p p_{m+1}}\right)
\,Z_{(1,2)}(u,v\,|\,p)=0,\ \ \ \ \ m\ge 0
\ }
\label{Vircon}
\ee
where
\be
\hat L_m^{\bf C}=\sum_{k=1}^\infty(m+k)p_k\frac{\p}{\p p_{m+k}}
+ \sum_{a=1}^{n-1} a(n-a)\frac{\p^2}{\p p_a\p p_{m-a}}
+ (u+v)\,m\frac{\p}{\p p_m}+ uv\delta_{m,0}
\label{Vir}
\ee
One can easily check that these constraints are indeed satisfied by (\ref{Zotau}) at $k=1$,
$n=2$.
Note that integration domain $x>0$ is  preserved by the transformation $\delta x = x^{m+1}$
only for $m\geq 0$, thus {\bf there is  no $\hat L_{-1}^{\bf C}$ constraint} --
this seems not to match the claim of \cite{AC}.
Let us stress that in case of (\ref{Vircon}) the second term in the brackets can be interpreted as the shift
of the $p_1$-variable,
but this is no longer so for more general $\widetilde W$-constraints,
see (\ref{lincon}) and \cite{Virmult,GKMU}.
Note also that we do not include $\p/\p p_0$ terms in the sum, and give the
corresponding contributions explicitly.
Usually they would be proportional to the matrix size $N$,
but in Virasoro constraints this size does not need to be integer.
Moreover, the would be $N^2$ is substituted by $uv$, while $2N$ by $(u+v)$.

\bigskip

Likewise, the $Z_{(1,3)}$ function (\ref{Z13}) satisfies the $\widetilde W^{(3)}$ constraint:
\be
\boxed{\
\left(\hat M_m^{\bf C}- \frac{m+1}{s}\frac{\p}{\p p_{m+1}}\right)
\,Z_{(1,3)}(u,v,w\,|\,p)=0,\ \ \ \ \ m\ge 0
\ }
\ee
where
\be
\hat M^{\bf C}_0 =
\sum_{a,b=1}^{\infty} \left((a+b)p_ap_b\frac{\p}{\p p_{a+b}} + abp_{a+b}\frac{\p^2}{\p p_a\p p_b}\right)
 +(u+v+w)\sum_{a=1}^\infty ap_a\frac{\p}{\p p_a} + uvw
\ee
and, more generally,
\be
\hat M^{\bf C}_m = \sum_{k,l=1}^\infty (k+l+m)p_kp_l\frac{\p}{\p p_{k+l+m}} +
\sum_{k=1}^\infty \Big(\sum_{a=1}^{k+m-1}+\sum_{a=1}^m\Big)\
a(k+m-a) p_k \frac{\p^2}{\p p_a\p p_{k+m-a}}
+  \\
+ \sum_{a+b+c=m} abc\frac{\p^3}{\p p_a\p p_b\p p_c} + uvw\delta_{m,0}
+ \frac{m^2(m+1)}{2}\frac{\p}{\p p_m}+(uv+vw+wu)m\frac{\p}{\p p_m} +  \\
+ (u+v+w)\left(\sum_{k=1}^\infty (k+m)p_k\frac{\p}{\p_{k+m}}
+\sum_{a+b=m} \frac{\p^2}{\p p_a\p p_b}\right)
\label{Wc}
\ee
Clearly, this time
$N^3 \longrightarrow uvw$, $3N^2\longrightarrow (uv+vw+wu)$ and $3N\longrightarrow (u+v+w)$.
We keep the same label ${\bf C}$ for these operators, to emphasize similarity with (\ref{Vir}).
In fact they belong to the class of the $\widetilde W$-operators \cite{Virmult,GKMU,UFN23},
appearing in description of Kontsevich and multi-matrix models and mnemonically are
powers of the current $\hat J^{\bf C}$ defined by (\ref{compcur}), subjected to peculiar normal ordering,
when all the $\hat J_-^{\bf C}$ operators on the right are simply thrown away, see \cite{Virmult}
for a detailed description.

\bigskip

Similarly, one can treat the models $Z_{(1,n)}$ with higher $n>3$. They satisfy similar $\widetilde W^{(n)}$-constraints.
In principle, they can be derived either from multi-matrix models or from any of the
$W$-representations, described in the present paper.

For illustrative purposes we provide just one more example:
\be
\left(\hat N_m^{\bf C}- \frac{m+1}{s}\frac{\p}{\p p_{m+1}}\right)
\,Z_{(1,4)}(u,v,w,x\,|\,p)=0,\ \ \ \ \ m\ge 0
\ee
and the simplest of operators $\widetilde W^{(4)}$ is
\be
\hat N^{\bf C}_0
=\sum_{a,b,c=1}^{\infty} \left((a+b+c)p_ap_bp_c\frac{\p}{\p p_{a+b+c}}
+ abcp_{a+b+c}\frac{\p^2}{\p p_a\p p_b\p p_c}\right)+\\
+ \frac{3}{2}\!\!\!\sum_{a+b=c+d} cd p_ap_b\frac{\p^2}{\p p_c\p p_d} +{1\over 2}
 \sum_{a,b=1}^\infty abp_ap_b \frac{\p^2}{\p p_a\p p_b}+\\
+\big(u+v+w+x\big)\sum_{a,b=1}^\infty \left((a+b)p_ap_b\frac{\p}{\p p_{a+b}}
+ abp_{a+b} \frac{\p^2}{\p p_a\p p_b}\right)
+ \big(uv+uw+ux+vw+vx+wx\big) \sum_{k=1}^\infty kp_k\frac{\p}{\p p_k}  +
\\
+ \sum_{k=1}^\infty \frac{k^2(k+1)}{2}p_k\frac{\p}{\p p_k} + uvwx
\ee

\section{Naive $W$-representations\label{glinfty}}

In addition to $W$-representation (\ref{OCas}) in terms of the Casimir operators,
which immediately implies integrability, one can rewrite the generating functions (\ref{WrepO}) as an exponential
in a more straightforward way, which also provides  nice expressions manifestly
belonging to integrability-preserving $GL(\infty)$ group \cite{DJKM,Kaz2}.

\subsection{The case of $Z_{(1,1)}$}

From (\ref{WrepO}) and from the fact that the operator $\hat O(u)$ in (\ref{OvsW})
preserves unity, $\hat O(u)\cdot 1 = 1$, it follows that
\be
\boxed{
Z_{(1,1)}(s,u)= \hat O(u)\,\circ e^{sp_1} \cdot 1=
\exp \Big( \hat O(u) \circ sp_1 \circ \hat O(u)^{-1}\Big)\cdot 1
}
\label{altW1}
\ee
(the last equality holds for any function, not obligatory exponential,
but $Z_{(1,1)}(s,u)$ is expressed via exponential).
Note that to use these kind of formulas one needs to rewrite
(\ref{Wef}) and (\ref{WrepO}) as some operator relations using composition $\circ$
instead of action of operators, i.e. $e^{sp_1}$ in (\ref{altW1}) is treated not as a function, but as an operator (of multiplication by $e^{sp_1}$).
For example, for $\hat W_{[1]} = \hat L_0 = \sum_nnp_n\frac{\p}{\p p_n}$ and
$\chi_{[1]} = p_1$ one has
\be
\hat W_{[1]}\circ\chi_{[1]}= \chi_{[1]} + \chi_{[1]}\circ\hat W_{[1]}
\ee
and (\ref{Wef}) is reproduced if we apply this identity to unity,
which is annihilated by $\hat W_\Delta$:
\be
\hat W_{[1]}\circ\chi_{[1]}\cdot 1 = \chi_{[1]}\cdot 1 + \chi_{[1]}\circ\hat W_{[1]}\cdot 1 =
\chi_{[1]}\cdot 1 = p_1
\ee
For the sake of brevity, we omit the sign of composition $\circ$ throughout this section,
since it is implied at any operator expressions here.

We can now use (\ref{OvsW}) to calculate the operator
$ \hat O(u) sp_1 \hat O(u)^{-1}$,which stands in the exponent in (\ref{altW1}).
For this we need the explicit formulas for $\hat W_\Delta$ from \cite{MMN1}.
For $\Delta = \delta_1\geq\delta_2\geq\ldots\geq\delta_{l(\Delta)}\geq 0
= \{\ldots, \underbrace{2,\ldots,2}_{m_2},\underbrace{1,\ldots,1}_{m_1}\}$
\be
\hat W_\Delta =   \prod_k
\frac{1}{m_k! k^{m_k}} :\hat D_k^{m_k} :
\ee
where $\hat D$ are defined in terms of the Miwa matrix $X$ from $p_k = \Tr X^k$:
\be
\hat D_k = \Tr \left(X\frac{\p}{\p X^{tr}}\right)^k = \Tr (X\p_X)^k
\ee
and the double dots denote normal ordering: all the $X$-derivatives
stand to the right of all $X$'s, e.g.
\be
:\Tr (X\p_X)^2:\ = \ :\,X_{ij}\frac{\p}{\p X_{kj}} X_{kl}\frac{\p}{\p X_{il}}\,:\ =
X_{ij}X_{kl} \frac{\p^2}{\p X_{kj}X_{il}}
\ee
(this example illustrates also the meaning of the transposition superscript $X^{tr}$).
It is because of the normal ordering that $\hat W_\Delta$ annihilates unity.

Now we can act with $\hat W_\Delta$ on $p_1$.
The commutator
\be
[:\hat D_k:\,,  p_1] = \ k:\Tr X^2\p_X (X\p_X)^{k-2}:
\ee
This implies that
\be
\hat W_2 \, p_1 = \frac{1}{2} :\hat D_2:\ p_1 = p_1 \hat W_2 + \underline{\Tr X^2\p_X}   \\
\hat W_3 \, p_1 = \frac{1}{3} :\hat D_3:\ p_1 = p_1 \hat W_3 + :\Tr X^2\p_X X\p_X:   \\
\hat W_{22}\, p_1 = \frac{1}{8} :\left(\hat D_2\right)^2:\ p_1=
p_1\hat W_{22} + \frac{1}{2}:\hat D_2\Tr X^2\p_X:
 \\
\ldots
\label{Wpcom}
\ee
Now, add the two last lines:
\be
\left[\big(\hat W_3+\hat W_{22}\big), p_1\right] = \frac{1}{2} \left(:\hat D_2\Tr X^2\p_X:
\ + \ 2:\Tr X^2\p_X X\p_X:\right) = \frac{1}{2} \Tr X^2\p_X :\hat D_2:
= \Big(\underline{\Tr X^2\p_X}\Big)\, \hat W_2
\ee
where the underlined operator is just the same as in the first line of (\ref{Wpcom}).

Coming back to (\ref{altW1}), we see that
\be
\hat O(u)p_1 = \left(1+ \frac{\hat W_2}{u} + \frac{\hat W_3+\hat W_{22}}{u^2} + \ldots \right)
u^{\hat L_0} p_1
= \left(u+ \hat W_2 + \frac{\hat W_3+\hat W_{22}}{u} + \ldots \right)p_1u^{\hat L_0} =  \\
= \left\{p_1\left(u+ \hat W_2 + \frac{\hat W_3+\hat W_{22}}{u} + \ldots \right)
+  \underline{\Tr X^2\p_X} + \underline{\Tr X^2\p_X}\,\frac{\hat W_2}{u} + \ldots \right\} u^{\hat L_0}
=  \\
= up_1 \hat O(u)+\Big(\underline{\Tr X^2\p_X}\Big)\,  \hat O(u) = (up_1+\hat L_{-1})\hat O(u)
\ \ \ \ \ \ \ \ \ \ \ \
\label{Oup}
\ee
where
\be
\hat L_0 = \hat W_{[1]} = \Tr X\p_X = \sum_m mp_m\frac{\p}{\p p_m},  \\
\hat L_{-1} = \underline{\Tr X^2\p_X} = \sum_m mp_{m+1}\frac{\p}{\p p_m}
\ee

Thus we obtain from (\ref{altW1}) a $W$-representation
\be
\boxed{
Z_{(1,1)}(s,u|p) = e^{s(\hat L_{-1} + up_1)}\cdot 1
}
\label{altZ2}
\ee
alternative to (\ref{OCas}).

\subsection{Direct check of (\ref{altZ2})}

In fact, $Z_{(1,1)}(s,u|p)$ is known explicitly,  see (\ref{Z2}).
The relation
\be
e^{s(\hat L_{-1} + up_1)}\cdot 1 = \exp\left(u\sum_m \frac{s^mp_m}{m}\right) = Z_{(1,1)}(s,u|p) =
\sum_R s^{|R|}D_R(u)\chi_R\{p\}
\label{Wrep1}
\ee
implied by (\ref{altZ2}),
follows from the Campbell-Hausdorff formula, if it is written in the form
\be
\exp\left(\frac{[B,A]}{2} - \frac{\left[A,[A,B]\right]}{3} + \frac{\left[[A,B],B\right]}{6} + \ldots
\right)\cdot e^A\cdot e^B = e^{A+B}
\label{CH1}
\ee
We choose $A=sup_1$ and $B = s\hat L_{-1}$, since in this case $e^B\cdot 1=1$.
Then only the first and the third terms at the very left exponential
contributes giving $us^2p_2/2$ and $us^3p_3/3$. More generally, only the terms of the form
\be
\sum_m \frac{{\rm ad}_B^m(A)}{m(m+1)}
\ee
contribute. Since clearly ${\rm ad}^m_B(A) = mp_{m+1}$, while all other commutators
(like $\sum_m \frac{{\rm ad}_A^m(B)}{m+1}$)
are vanishing,
\be
\!\!\!\!\!\!\!\!\!\!\!\!
e^{s(\hat L_{-1} + up_1)}\cdot 1 = e^{A+B}\cdot 1 =
\exp\left(\sum_{m=1} \frac{{\rm ad}_B^m(A)}{m(m+1)}\right) e^A
= \exp\left(\sum_{m=1} \frac{s^{m+1}up_{m+1}}{m+1}\right) \cdot e^{sup_1}
= \exp\left(u\sum_{m=1} \frac{s^mp_m}{m}\right)
\ee
which is exactly (\ref{Wrep1}).

\subsection{The case of $Z_{(1,2)}$}

This time instead of (\ref{altW1}) one needs
\be
Z_{(1,2)}(s,u,v)= \hat O(v)\hat O(u) \, e^{sp_1} \cdot 1=
\exp \Big( \hat O(v)\hat O(u) sp_1 \hat O(u)^{-1}\hat O(v)^{-1}\Big)\cdot 1
\label{altW2}
\ee
and thus an appropriate modification of (\ref{Oup}):
\be
\hat O(v)\hat O(u)p_1  = \hat O(v)\left(
up_1 \hat O(u)+\Big(\underline{\Tr X^2\p_X}\Big)\,  \hat O(u)\right) =  \\
= uvp_1\hat O(v)\hat O(u) + u \Big(\underline{\Tr X^2\p_X}\Big)\, \hat O(v) \hat O(u)
+ \hat O(v) \Big(\underline{\Tr X^2\p_X}\Big)\,  \hat O(u) =  \\
= uvp_1\hat O(v)\hat O(u) + (u+v) \Big(\underline{\Tr X^2\p_X}\Big)\, \hat O(v) \hat O(u)
- \left[\underline{\Tr X^2\p_X}\,, \
 \left(v+ \hat W_2+ \frac{\hat W_3+\hat W_{22}}{v}+\ldots\right)\right] \hat O(u) =  \\
= uvp_1\hat O(v)\hat O(u) + (u+v) \Big(\underline{\Tr X^2\p_X}\Big)\, \hat O(v) \hat O(u)
+ \Big(\underline{\underline{:\Tr X^2\p_X X\p_X:}}\Big)\, \hat O(v) \hat O(u) = \\
= \Big(uvp_1+(u+v)\hat L_{-1} + \hat M_{-1}\Big) \hat O(v) \hat O(u)
\ \ \ \ \ \ \ \
\ee
with
\be
\hat L_{-1} = \underline{\Tr X^2\p_X} = \sum_m mp_{m+1}\frac{\p}{\p p_m},  \\
\hat M_{-1} = \underline{\underline{:\Tr X^2\p_X X\p_X:}}
= \sum_{a,b} (a+b-1)p_ap_b \frac{\p}{\p p_{a+b-1}} + ab p_{a+b+1}\frac{\p^2}{\p p_a\p_b}
\ee
Combining this with (\ref{altW2}) we immediately reproduce the result of \cite{Zog}:
\be\label{z12}
\boxed{
Z_{(1,2)}(s,u,v)= \exp\left\{s\Big(uvp_1+(u+v)\hat L_{-1} + \hat M_{-1}\Big)\right\}\cdot 1
}
\ee

\subsection{Operators $\hat {\cal O}(u_1,\ldots,u_n)$}

Now generalizing (\ref{altW2}), one can define the operator $\hat {\cal O}(u_1,\ldots,u_n)$
\be
Z_{(1,n)}(s,u_1,\ldots, u_n)=
\exp \Big( \prod_{i=1}^k\hat O(u_i) sp_1 \prod_{I=1}^n\hat O(u_i)^{-1}\Big)
\cdot 1=\hat {\cal O}(u_1,\ldots,u_n)\cdot 1
\label{altW2n}
\ee
The sequence of underlined operators is evidently
\be
{\rm ad}_{\hat W_2}^k p_1 = \,:\Tr X^2\p_X(X\p_X)^{k-1}:
\ee
in particular,
\be
\hat L_{-1} = [\hat W_2\,,\ p_1] = \Tr X^2\p_X,  \\
\hat M_{-1} = [\hat W_2\,,\ \Tr X^2\p_X] = \,:\Tr X^2\p_XX\p_X:   \\
\hat N_{-1} = [\hat W_2\,, \ :\Tr X^2\p_X X\p_X:] =\,:\Tr X^2\p_X(X\p_X)^2:   \\
\ldots
\ee
Therefore the naive $W$-representations of the functions $Z_k$ look as follows:
\be\boxed{
Z_{(1,k)}(\vec u) = \hat {\cal O}_k(\vec u)\cdot 1}
\ee
where
\fr{
\hat {\cal O}_1 = e^{sp_1},  \\
\hat {\cal O}_2(u) = e^{s(\hat L_{-1} + up_1)}, \\
\hat {\cal O}_3(u,v) \ \stackrel{\cite{Zog}}{=}
e^{s(\hat M_{-1} + (u+v)\hat L_{-1} + uvp_1)},  \\
\hat {\cal O}_4(u,v,w) = e^{s(\hat N_{-1} + (u+v+w)\hat M_{-1} + (uv+vw+wu)\hat L_{-1} + uvwp_1)},  \\
\ldots
\label{calOk}}
and
\fr{
\hat L_{-1} = \sum_m   mp_{m+1}\frac{\p}{\p p_m},  \\
\hat M_{-1} = \sum_{a,b} (a+b-1)p_ap_b \frac{\p}{\p p_{a+b-1}}
+ ab p_{a+b+1}\frac{\p^2}{\p p_a\p_b},  \\
\hat N_{-1} =\sum_{a,b,c=1}^\infty \left((a+b+c-1)\,p_ap_bp_c \,\frac{\p\ \ \ \ \ }{\p p_{a+b+c-1}}
+ abc\, p_{a+b+c+1}\,
\frac{\p^3\ \  }{\p p_a\p p_b\p p_c} \right)+  \\
+ \ \frac{3}{2} \sum_{a,b=1}^\infty \,\sum_{c=1}^{a+b} ab\, p_cp_{a+b+1-c}\,\frac{\p^2}{\p p_a\p p_b}
+\ \frac{1}{2}\sum_{a=1}^\infty a^2(a+1)\,p_{a+1}\,\frac{\p}{\p p_a},  \\
\ldots
}
Formula (\ref{z12}) for $Z_{(1,2)}$ appeared in \cite{Zog}.

Note that this representation of the
operators $\hat {\cal O}_k(\vec u)$ also makes manifest that they are elements of $GL(\infty)$
\cite{DJKM,Kaz2} which gives yet another proof of integrability: this property
guarantees that $Z_{(1,n)}(\vec u)$ is a $\tau$-function of the KP hierarchy.

\subsection{Hierarchy in $n$ }

Operators (\ref{calOk}) form a clear hierarchy in $n$,
and one can easily move in $n$
in both directions.
Let us look at the simpler one: the decrease of $n$.

Since $D_R(v) = d_Rv^{|R|}\Big(1+O(v^{-1})\Big)$, one has
\be
\lim_{v\longrightarrow\infty} Z_{(1,n+1)}\left(\frac{s}{v},\vec u,v\right) = Z_{(1,n)}(s,\vec u)
\ee
For example, for $n=0$,
\be
\lim_{v\longrightarrow \infty}  \exp\left(\sum_m \frac{(s/v)^m\cdot v}{m}p_m\right)=
e^{sp_1}
\ee

Thus
\be
\hat{\cal O}_n(\vec u) = \lim_{v\longrightarrow \infty}\hat{\cal O}_{n+1}(v,\vec u)^{1/v}
\ee
In particular, taking $\hat{\cal O}_2$ from \cite{Zog},
we immediately get:
\be
\ldots \longrightarrow
\exp\left\{s\Big(\hat M_{-1} + (u+v)\hat L_{-1} + uvp_1\Big)\right\}
\longrightarrow \exp\left\{s(\hat L_{-1} + up_1)\right\}
\longrightarrow e^{sp_1}
\ee
It now looks rather obvious that the previous term on the left is
\be
\exp\left\{s\Big(\hat N_{-1}+(u+v+w)\hat M_{-1} +(uv+vw+wu)\hat L_{-1} +  uvwp_1\Big)\right\}
\ee
and so on.

\section{Description in terms of the $w_\infty$-algebra
\label{A}}

The $W$-representation (\ref{OCas}) can be further transformed and simplified.
Since it is expressed through the Casimir operators (\ref{staCas}),
which belong to the $W_\infty$ algebra,
and no central extensions are relevant for our considerations,
one can make use of its alternative representation in terms of
ordinary differential operators \cite{FKN2}.
This is a very powerful technique, see \cite{Alex} for the recent review,
and this also turns to be the case in application to our problem.

\subsection{Combined Casimir operators $\hat{\widetilde {C}}$ as distinguished $\hat W^{(m)}_0$}

In this approach
operators from $w_\infty$ are represented by polynomial of $z$ and $D=z\p_z$.
In most considerations $D$ can be considered just as an integer number.
In particular, the standard Casimir operators (\ref{staCas}) are mapped \cite{FKN2,Alex} into
\be
\hat C(n) \ \ \longrightarrow \ \ \frac{\left(D-\frac{1}{2}\right)^n - \left(-\frac{1}{2}\right)^n}{n}
\ee
Substituting this into the sums in (\ref{sigmatilde}), we obtain that
combined Casimir operators, given by this seemingly complicated formula,
are in fact mapped into something clearly distinguished:
\be
\hat{\widetilde C}(n+1) \ \ \longrightarrow \ \  \sum_{i=1}^{D-1} i^n
\label{tildeCD}
\ee
and then, from (\ref{OCas})
\be
\boxed{
{\hat O}(u) = u^{\hat C_1}\exp \left\{  \sum_{n=1}^\infty
\frac{(-)^{n+1}}{u^n \cdot n}\, \hat{\widetilde C}(n+1) \right\}
\ \ \longrightarrow \ \
u^D\exp\left(\sum_{i=0}^{D-1}\log\Big(1+\frac{i}{u}\Big)\right)
= \frac{\Gamma(u+D)}{\Gamma(u)}
}
\label{OCasGamma}
\ee
i.e. as an element of the $w_\infty$ algebra, {\bf operator $\hat O(u)$
is just an ordinary $\Gamma$-function}! \
In fact, Bernoulli numbers naturally arise in the coefficients of the
large-$u$ asymptotics of $\log \Gamma(u)$.

Moreover, the sums at the r.h.s. (\ref{tildeCD}) are also associated with the very
special operators, what provides a spectacular interpretation of $\hat{\widetilde{C}}(n)$.
Namely, monomials $z D^n$ are images of
\be
\begin{array}{cc}
p_1 \ \ &\longrightarrow \ \ z\cdot 1,  \\
\hat L_{-1} = \sum_n np_{n+1}\frac{\p}{\p p_n} \ \ &\longrightarrow \ \ z\cdot D,  \\
\hat M_{-1} = \sum_{a,b}
\left((a+b-1)p_ap_b\frac{\p}{\p p_{a+b-1}} + abp_{a+b+1}\frac{\p^2}{\p p_a\p p_b}\right)
\ \ &\longrightarrow \ \ z\cdot D^2,  \\
\ldots
\label{min1vsD}
\end{array}
\ee
and the sums in (\ref{tildeCD}) are the zeroth harmonics of the same operators:
\be
\begin{array}{ccc}
\hat L_0 = \sum_n np_{n+1}\frac{\p}{\p p_n} \ \ &\longrightarrow \ \ D &= \sum_{i=1}^{D-1} 1,  \\
\hat M_{0} = \sum_{a,b}
\left((a+b)p_ap_b\frac{\p}{\p p_{a+b}} + abp_{a+b}\frac{\p^2}{\p p_a\p p_b}\right)
\ \ &\longrightarrow \ \ D(D-1) &= 2\sum_{i=1}^{D-1} i,  \\
\hat N_0 =  \sum_{a,b,c=1}^\infty \left((a+b+c)\,p_ap_bp_c \,\frac{\p\ \ \ \ \ }{\p p_{a+b+c}}
+ abc\, p_{a+b+c}\,
\frac{\p^3\ \  }{\p p_a\p p_b\p p_c} \right)+
\!\!\!\!\!\!\!\!\!\!\!\!\!\!\!\!\!\!\!\!\!\!\!\!  \\
+ \ \frac{3}{2} \sum_{a,b=1}^\infty \,\sum_{c=1}^{a+b-1} ab\, p_cp_{a+b-c}\,\frac{\p^2}{\p p_a\p p_b}
+\ \frac{1}{2}\sum_{a=1}^\infty a(a^2-1)\,p_{a}\,\frac{\p}{\p p_a}\ \
&\longrightarrow \ \ \frac{1}{2}D(D-1)(2D-1) &= 3\sum_{i=1}^{D-1} i^2  \\
\ldots
\end{array}
\label{min0vsD}
\ee
Let us introduce a unified notation  $\W^{(m)}_n$ for all these $W$-operators:
\be
p_k = \W^{(1)}_k, \ \ \hat L_k = \W^{(2)}_k, \ \ \hat M_k = \W^{(3)}_k, \ \
\hat N_k = \W^{(4)}_k, \ldots
\ee
Comparing (\ref{tildeCD}) with (\ref{min0vsD}) we see that
\be
\boxed{
\hat{\widetilde C}(n) = \frac{1}{n} \hat W^{(n+1)}_0
}
\ee
In terms of these operators one can rewrite (\ref{OCas}) and (\ref{OCasGamma}) as
\be
\boxed{
\
\hat O(u) =
\exp\Big(\log u\ \hat L_0 + \frac{1}{2u}\hat M_0 - \frac{1}{6u^2}\hat N_0 + \ldots\Big)
= \exp\left(\sum_{m=2}^\infty  \frac{(-)^m\W^{(m+1)}_0}{(m-1)m\, u^{m-1}}\right)
u^{\hat W^{(2)}_0}
\
}
\label{prealtW}
\ee
so that
\be
Z_{(1,n)}(s,\vec u) = \prod_{i=1}^n \hat O(u_i)\cdot e^{sp_1}
= \exp\left(\sum_{m=2}^\infty \eta_m  \W^{(m+1)}_0\right)\cdot
\exp\left(sp_1\prod_{i=1}^n u_i\right)
\label{ZW0}
\ee
with
\be
\eta_m = \frac{(-)^{m}}{(m-1)m}\sum_{i=1}^n \frac{1}{u_i^{m-1}}
\ee

\subsection{Relation between the two $W$-representations}

At the same time, from (\ref{calOk}) the same function is given by
\be
Z_{1,n}(s,\vec u)
=\exp \left(su_1\ldots u_n\left\{
p_1 + \sum_{i=1}^n \frac{1}{u_i}\, \hat L_{-1} +
\sum_{i<j}^n \frac{1}{u_iu_j}\, \hat M_{-1} +
\sum_{i<j<k}^n \frac{1}{u_iu_ju_k}\,\hat N_{-1} +
\ldots
\right\}\right) \cdot 1 =  \\
=\exp \left(s\Big(\prod_{i=1}^n u_i\Big)\Big(\sum_{m=0}^\infty \xi_m \W_{-1}^{(m+1)}\Big)\right)\cdot 1
\label{ZW1}
\ee
with
\be
\xi_m = \sum_{i_1\leq i_2\leq\ldots\leq i_m} \frac{1}{u_{i_1}u_{i_2}\ldots u_{i_m}}
\ee
In this form there are two differences between (\ref{ZW0}) and (\ref{ZW1}):
the grading of $\W$-operators ($0$ and $-1$ respectively)
and the time variables $\eta$ and $\xi$, given respectively
by power sum and elementary symmetric
polynomials of  variables $u_i^{-1}$.

These two $W$-representations are of course related by the Campbell-Hausdorff formula,
this time in the form
\be
e^{\hat B} e^{\hat A} = \boxed{e^{\hat A + [\hat B,\hat A] + \frac{1}{2!}[\hat B[\hat B,\hat A]]
+ \ldots }}\ e^{\hat B}
\label{CHf}
\ee
when exponent in the boxed operator is just
\be
\hat C = \sum_{m=0}^\infty \frac{1}{m!} {\rm ad}_{\hat B}^m\hat A
\label{CvsAB}
\ee
where we need to substitute $\hat A = p_1$ and $\hat B = \sum_m \eta_m \W^{(m)}_0$.
Since (\ref{CvsAB}) is linear in $\hat A$, the common factor $s\prod u_i$ can be
omitted and restored at the very end.
Then, if applied to unity, the l.h.s. of (\ref{CHf}) gives (\ref{ZW0}), and the r.h.s.
will provide (\ref{ZW1}), because $e^{\hat B}\cdot 1 = 1$.
To calculate $\hat C$ we need a commutation relation
\be
\big[ \W_0^{(m+1)}, \, \W_{-1}^{(n+1)}\big] = m\W_{-1}^{m+n}
\ee
which provides $\hat C$ in the following form:
\be
\hat C = \underbrace{\W^{(1}_{-1}}_{p_1} + \sum_{m=2}^\infty m\,\eta_m \W^{(m)}_{-1}
+ \frac{1}{2!}\sum_{m,n=2}^\infty mn\,\eta_m\eta_n \W^{(m+n)}_{-1}
+ \frac{1}{2!}\sum_{l,m,n=2}^\infty lmn\,\eta_l\eta_m\eta_n \W^{(l+m+n)}_{-1} + \ldots
\ee
We want this to be equal to
$\ \ \sum_{k=0}^\infty \xi_k \W^{(k+1)}_{-1}\ \ $
Clearly, each $\xi_k$ is a {\it finite} multi-linear combination of $\eta_m$,
for example,
\be
\xi_0 = 1,  \\
\xi_1 = 2\eta_2 = \sum_i \frac{1}{u_i},  \\
\xi_2 = 3\eta_3 + 2\eta_2^2 = -\frac{1}{2}\sum_i \frac{1}{u_i^2} + \frac{1}{2}
\left(\sum_i \frac{1}{u_i}\right)^2 = \sum_{i<j} \frac{1}{u_iu_j},  \\
\xi_3 = 4\eta_4+6\eta_2\eta_3+\frac{4}{3}\eta_2^2 = \sum_{i<j<k} \frac{1}{u_iu_ju_k},  \\
\ldots
\ee
Thus (\ref{ZW0}) and (\ref{ZW1}) -- and thus (\ref{OCas})  and (\ref{calOk}) --
are indeed related by the simplest of all Campbell-Hausdorff formulas (\ref{CHf}).

\subsection{More details from the $w_\infty$ dictionary}

Higher harmonics of the simplest operators $\W^{(m)}$ are mapped into
the following polynomials of $z$ and $D=z\p_z$:
\be\label{obdif}
\hat{J}_k=\mbox{res}_z(z^{k}\hat{J}(z)) \ \longrightarrow\ j_k=z^{-k},\,\,\,\,\,\,\,\, k\neq 1,\\
\hat{L}_k=\frac{1}{2}\mbox{res}_z\left(z^{1+k}:\hat{J}(z)^2:\right)
\ \longrightarrow\ l_k=z^{-k}\left(z \p_z-\frac{k+1}{2}\right),\\
\hat{M}_k=\frac{1}{3}\mbox{res}_z\left(z^{2+k}:\hat{J}(z)^3:\right)
\ \longrightarrow\ m_k=z^{-k}\left(z^2 \p_z^2-k z \p_z +\frac{(1+k)(2+k)}{6}\right),\\
\hat N_0 \ \longrightarrow\ \frac{1}{2}(2z\p_z-1)(z\p_z-1)z\p_z, \ \ \ \ \ \ \
\hat N_{-1}\ \longrightarrow\ z(z\p_z)^3
\ee
(polynomials at the r.h.s. are defined up to constant terms, which do not affect commutators
-- expressions in (\ref{min0vsD}) make use of this freedom).
In general, for peculiar operators, which are made from the current
\be
\hat J(x) = \sum_m \frac{\hat J_m}{x^{m+1}} =
\sum_{m=1}^\infty \left(p_m x^{m-1} + \frac{m}{x^{m+1}}\frac{\p}{\p p_m}\right)
\label{Jw}
\ee
and its derivatives -- and at the same time belong to the $W_\infty$ algebra --
the mapping rule is:
\be
\boxed{
\mbox{res}_z\left(z^{-k}: \frac{(\hat{J}(z)+\p_z)^{m+1}}{m+1} :\ 1\right)
\ \longrightarrow\  \left(z^2\p_z \right)^m z^{k}
}
\label{opw}
\ee
It is easy to check that above examples fit into this scheme, with
\be
\hat L(x) = \sum_{m} \frac{\hat L_m}{x^{m+2}} = \ :\hat J(x)^2:  \\
\hat M(x) = \sum_m \frac{\hat M_m}{x^{m+3}} = \ :\hat J(x)^3:  \\
\hat N(x) = \sum_m \frac{\hat N_m}{x^{m+4}}= \ :\hat J(x)^4 - \big(\p_x \hat J(x)\big)^2:  \\
\ldots
\ee

Note, that this formalism is applicable only to operators from $W_\infty$ algebra,
i.e. those made from the current (\ref{Jw}) and its derivatives in a very special way --
as linear combinations of those at the l.h.s. of (\ref{opw}).
Already the forth power of the current, $:\hat J^4:$, does {\it not} belong to this algebra --
this is the reason for the $(\p \hat J)^2$ subtraction in $\hat N \in W_\infty$.
Another typical example are Virasoro operators $\hat L^{\bf C}_n$ in (\ref{Vir}).
They are actually made from the square of {\it another} current,
\be
\hat J^{\bf C}(x) = \sum_{m=1}^\infty \left(\frac{1}{2}p_m x^{m-1}
+ \frac{m}{x^{m+1}}\frac{\p}{\p p_m}\right)
\label{compcur}
\ee
with additional factor $1/2$ in the poshtive harmonics.
Because of this the $w_\infty$ technique, described in this section, can {\it not}
be used to prove and even check the Virasoro constraints (\ref{Vir}):
it does {\it not} adequately describe commutation relations between
$\hat L^{\bf C}_n \notin W_\infty$ and $\hat L_0,\hat M_0,\hat N_0,\ldots \in W_\infty$.
However, there are two amusing exceptions: the zero harmonics $\hat L^{\bf C}_0$ and
$\hat M^{\bf C}_0$ do belong to $W_\infty$, this is no longer true neither for
$\hat N^{\bf C}_0$, nor for higher harmonics of $\hat L^{\bf C}$ and $\hat M^{\bf C}$.

\section{Conclusion}

This paper gives a brief summary of existing knowledge about the simple family (\ref{Zotau})
with $k=1,2$.
This family consists of Hurwitz $\tau$-functions
which are integrable in the simplest KP/Toda sense.
A number of facts are already present in the literature, not only we presented them in a systematic way
revealing all the relations between these facts, but we naturally made a number of new claims:
\begin{itemize}
\item In addition to the naive $W$-representation in s.\ref{glinfty} we described two others:
in terms of the generalized cut-and-join operators, (\ref{OvsW})
and of the Casimir operators, (\ref{OCas}), providing a direct relation to the Hurwitz theory
{\it a la} \cite{MMN1} and to the KP/Toda integrability
respectively.
One more version, (\ref{prealtW}), provides a bridge between naive and Casimir $W$-representations.
\item We put together the two-matrix and Kontsevich like models from \cite{OS,AC}
and pointed out an intriguing relation of higher $Z_{(2,n)}$ to
the conformal like matrix models.
\item We provided a description of the most studied $Z_{(1,2)}$ model
in terms of complex matrix model which directly provides the Virasoro constraints, (\ref{Vir}).
Similarly, the $Z_{(1,3)}$ model is described by the asymmetric two-matrix model with $1/Y$ potential
and satisfies the ${\widetilde W}^{(3)}$-constraints, etc.
\item We interpreted (-1)-modes of $W$-operators which enter the naive $W$-representation of \cite{Zog}
and its generalizations as multiple commutators of
the basic pair: the cut-and-join operator $\ \hat W_{[2]}={1\over 2}:\Tr (X\partial_X)^2:\ $ and
$\ \hat L_{-1}=\,:\Tr (X^2\partial_X):$
\item We explained in s.\ref{A} how the mapping to the differential operators can be used
to drastically simplify derivation of these and many other similar results
(note, however, that this approach is directly applicable only to the KP/Toda,
but not to general Hurwitz $\tau$-functions, and is thus restricted to models (\ref{Zotau})).
\end{itemize}
There are still a lot of formulas to derive, especially for  $Z_{(2,n)}$ models with $n>1$.

\section*{Acknowledgements}

We are grateful to L.Chekhov for useful discussions.
Our work is partly supported by ERC Starting Independent Researcher Grant StG No. 204757-TQFT (A.A.), the grants
NSh-1500.2014.2 (A.A., A.M.'s) and NSh-5138.2014.1 (S.N.),
by RFBR  13-02-00457 (A.A., A.Mir. and S.N.), 13-02-00478 (A.Mor.),
by joint grants 13-02-91371-ST, 14-01-92691-Ind,
by the Brazil National Counsel of Scientific and
Technological Development (A.Mor.),
by Laboratory of Quantum Topology of Chelyabinsk State University
(Russian Federation government grant 14.Z50.31.0020) (S.N.) and by FRIAS (A.M.'s).


\begin{thebibliography}{12}


\bibitem{GJ}
I.P.Goulden and D.M.Jackson, arXiv:0803.3980

\bibitem{Zog} P.Zograf,
arXiv:1312.2538

\bibitem{AC} J.Ambjorn and L.Chekhov, arXiv:1404.4240

\bibitem{Kaz} M.Kazaryan and P.Zograf, to appear

\bibitem{MMN1} A.Mironov, A.Morozov and S.Natanzon,
Theor.Math.Phys. {\bf 166} (2011) 1-22,
arXiv:0904.4227;  Journal of Geometry and Physics {\bf 62} (2012) 148-155,
arXiv:1012.0433

\bibitem{Hurtau} S.Kharchev, A.Marshakov, A.Mironov and A.Morozov, Int.J.Mod.Phys. {\bf A10} (1995) 2015,
hep-th/9312210\\
A.Alexandrov, A.Mironov, A.Morozov and S.Natanzon,
J.Phys. A: Math.Theor. {\bf 45} (2012) 045209,
arXiv:1103.4100

\bibitem{Slep} A.Mironov, A.Morozov and A.Sleptsov,
Theor.Math.Phys. 177 (2013) 1435-1470 (Teor.Mat.Fiz. 177 (2013) 179-221),  arXiv:1303.1015;
European Physical Journal C 73 (2013) 2492,  arXiv:1304.7499;
arXiv:1310.7622

\bibitem{OS}
A.Orlov and D.M.Shcherbin,
Theor.Math.Phys.
{\bf 128} (2001) 906-926\\
A.Orlov,
Theor.Math.Phys. {\bf 146}
(2006) 183–206


\bibitem{Dij} R.Dijkgraaf,
In: {\sl The moduli
spaces of curves},
Progress in Math., 129 (1995), 149-163,
Brikh\"auser

\bibitem{Mac} D.E.Littlewood, {\sl The theory of group characters and
matrix representations of groups}, Oxford, 1958\\
M.Hamermesh, {\sl  Group theory and its application to physical problems},
1989\\
I.G.Macdonald, {\sl Symmetric functions and Hall polynomials}, Oxford Science
Publications, 1995\\
W.Fulton, {\sl Young tableaux: with applications to representation theory and geometry},
London Mathematical Society, 1997

\bibitem{MMNWDVV}
A.Mironov, A.Morozov and S.Natanzon,
JHEP {\bf 11} (2011) 097, arXiv:1108.0885

\bibitem{UnInt} A.Morozov,
Teor.Mat.Fiz. {\bf 161} (2010) 3-40,
arXiv:0906.3518

\bibitem{Bel} G.Belyi, Mathematics of the USSR: Izvestiya, {\bf 14:2} (1980) 247-256\\
A.Grothendieck, {\sl Sketch of a Programme}, Lond. Math. Soc. Lect.
Note Ser. {\bf 242} (1997) 243-283;
{\sl Esquisse d'un Programme}, in: P.Lochak, L.Schneps (eds.), Geometric Galois Action, pp.5-48,
Cambridge University Press, Cambridge (1997)\\
G.B.Shabat and V.A.Voevodsky,
The Grothendieck Festschrift, Birkhauser,
1990, V.III., p.199-227\\
S.K.Lando and A.K.Zvonkin, {\sl Graphs on surfaces and their applications}, Encycl.
of Math. Sciences, {\bf 141}, Springer, 2004

\bibitem{BHM} C.Itzykson and J.B.Zuber, 
Commun. Math. Phys. {\bf 134} (1990) 197;\\
R. de Mello Koch and S.Ramgoolam, 
arXiv:1002.1634;\\
T.W.Brown, 
Phys.Rev. {\bf D83}  (2011) 085002, arXiv:1009.0674


\bibitem{Wit} E.Witten, 
Nucl.Phys. {\bf B340} (1990) 281-332

\bibitem{Kon}  M.Kontsevich, Funk.Anal. i Priloz. {\bf 25} (1991) 50

\bibitem{Pol} A.M.Polyakov, 
Phys. Lett. {\bf B103} (1981) 207-210; {\it ibid.}, pp.211-213

\bibitem{MMMeq} A.Marshakov, A.Mironov, A.Morozov, 
Phys.Lett., {\bf B274} (1992) 280, hep-th/9201011

\bibitem{Wit2} E.Witten, 
in: New York 1991 Proc., Differential geometric
methods in theoretical physics, v.1, pp.176-216

\bibitem{2dg} V.Kazakov, 
Mod.Phys.Lett. {\bf A4} (1989) 2125\\
E.Brezin and V.Kazakov, 
Phys.Lett. {\bf 236B} (1990) 144\\
M.Douglas and S.Shenker, 
Nucl.Phys. {\bf B335} (1990) 635\\
D.Gross and A.Migdal, 
Phys.Rev.Lett. {\bf 64} (1990) 127

\bibitem{D} M.Douglas, 
Phys.Lett. {\bf B238} (1990) 176

\bibitem{FKN} M.Fukuma, H.Kawai and R.Nakayama, 
Int.J.Mod.Phys. {\bf A6} (1991) 1385\\
R.Dijkgraaf, E.Verlinde and H.Verlinde, 
Nucl.Phys. {\bf B348} (1991) 435

\bibitem{LevMor} A.Levin and A.Morozov, 
Phys.Lett. {\bf B243} (1990) 207-214

\bibitem{Gop} R.Gopakumar, arXiv:1104.2386

\bibitem{Sha} N.Adrianov, N.Amburg, V.Dremov, Yu.Levitskaya, E.Kreines, Yu.Kochetkov, V.Nasretdinova, G.Shabat, arXiv:0710.2658

\bibitem{UFN23} A.Morozov,
Sov.Phys.Usp. \textbf{35} (1992) 671-714;
Sov.Phys.Usp. \textbf{37 }(1994) 1-55, hep-th/9303139;
hep-th/9303139; hep-th/9502091;
hep-th/0502010;\\
A.Mironov,
Int.J.Mod.Phys. {\bf A9} (1994) 4355, hep-th/9312212;
Phys.Part.Nucl.
{\bf 33} (2002) 537;
Theor.Math.Phys. \textbf{146} (2006) 63-72, hep-th/0506158

\bibitem{MorMaMo} A.Morozov,
 arXiv:1204.3953

 \bibitem{AMMEO} A.Alexandrov, A.Mironov and A.Morozov,
Int.J.Mod.Phys. {\bf A19} (2004) 4127, hep-th/0310113;
Theor.Math.Phys. {\bf 150} (2007) 153-164, hep-th/0605171;
Physica {\bf D235} (2007) 126-167, hep-th/0608228; JHEP {\bf 12} (2009) 053,
arXiv:0906.3305;\\
A.Alexandrov, A.Mironov, A.Morozov, P.Putrov,
Int.J.Mod.Phys. {\bf A24} (2009) 4939-4998, arXiv:0811.2825;\\
B.Eynard,
JHEP \textbf{0411} (2004) 031, hep-th/0407261;\\
L.Chekhov and B.Eynard,
JHEP \textbf{0603} (2006) 014, hep-th/0504116;
JHEP \textbf{0612} (2006) 026, math-ph/0604014;\\
N.Orantin,
arXiv:0808.0635

\bibitem{Wreps} A.Morozov and Sh.Shakirov,
 JHEP {\bf 0904} (2009) 064,  arXiv:0902.2627;
Mod.Phys.Lett. {\bf A24} (2009) 2659-2666,  arXiv:0906.2573\\
A.Alexandrov,
arXiv:1005.5715

\bibitem{Zhe} D.P.Zhelobenko, Compact Lie group and their representations,
American Mathematical Society, 1973

\bibitem{eqh} T.Shiota, Invent.Math. {\bf 83} (1986) 333\\
S.Kharchev, A.Marshakov, A.Mironov, A.Morozov,
Mod.Phys.Lett. {\bf A8} (1993) 1047-1061, hep-th/9208046\\
S.Kharchev, 
hep-th/9810091

\bibitem{Alex} A.Alexandrov, arXiv:1404.3402

\bibitem{DJKM} E.Date, M.Jimbo, M.Kashiwara, T.Miwa, {\it  Transformation groups for soliton
equations},
RIMS Symp. {\sl ``Non-linear integrable systems -- classical theory and
quantum
theory"} (World scientific, Singapore, 1983)

\bibitem{versus} S.Kharchev, A.Marshakov, A.Mironov and A.Morozov,
Nucl.Phys. \textbf{B397} (1993) 339-378, hep-th/9203043

\bibitem{Taka} K.Takasaki, Adv.Studies in Pure Math. {\bf 4} (1984) 139-163

\bibitem{KLMMM} A.Gerasimov, S.Khoroshkin, D.Lebedev, A.Mironov and A.Morozov,
Int.J.Mod.Phys. A10 (1995) 2589-2614, hep-th/9405011   \\
S.Kharchev, A.Mironov and A.Morozov,
q-alg/9501013;\\
A.Mironov, hep-th/9409190; Theor.Math.Phys. {\bf 114} (1998) 127, q-alg/9711006

\bibitem{Ok} A.Okounkov and R.Pandharipande,
Ann. of Math. {\bf 163} (2006) 517,
math.AG/0204305\\
S.Lando,
In: {\sl Applications of Group Theory to Combinatorics}, Koolen, Kwak and Xu, Eds.
Taylor \& Francis Group, London, 2008, 109-132

\bibitem{CJR} S.Corley, A.Jevicki and S.Ramgoolam, Adv.Theor.Math.Phys. {\bf 5} (2002) 809-839, hep-th/0111222

\bibitem{GKM} S.Kharchev, A.Marshakov, A.Mironov, A.Morozov and A.Zabrodin,
Phys. Lett. \textbf{B275} (1992) 311-314, hep-th/9111037;
Nucl.Phys. \textbf{B380} (1992) 181-240, hep-th/9201013\\
A.Mironov, A.Morozov and G.Semenoff, 
Int.J.Mod.Phys. {\bf A10} (1995) 2015, hep-th/9404005

\bibitem{confMaMo}
S.Kharchev, A.Marshakov, A.Mironov, A.Morozov and S.Pakuliak,
Nucl.Phys. {\bf B404} (1993) 717-750, hep-th/9208044\\
A.Mironov and S.Pakuliak,
Int.J.Mod.Phys. {\bf A8} (1993) 3107-3137, hep-th/9209100\\
H.Awata, Y.Matsuo, S.Odake and J.Shiraishi,
Soryushiron Kenkyu {\bf 91} (1995) A69-A75, hep-th/9503028

\bibitem{confdop}
 A.Marshakov, A.Mironov and A.Morozov,
Phys.Lett. {\bf B265} (1991) 99-107\\
R.Dijkgraaf and C.Vafa, arXiv:0909.2453;\\
H.Itoyama, K.Maruyoshi and T.Oota,
Prog.Theor.Phys. { 123} (2010) 957-987, arXiv:0911.4244;\\
T.Eguchi and K.Maruyoshi,
arXiv:0911.4797;
arXiv:1006.0828;\\
R.Schiappa and N.Wyllard,
arXiv:0911.5337;\\
A.Mironov, A.Morozov and Sh.Shakirov,
JHEP { 02} (2010) 030, arXiv:0911.5721;
Int.J.Mod.Phys. { A25} (2010) 3173-3207, arXiv:1001.0563;
Int.J.Mod.Phys. A27 (2012) 1230001, arXiv:1011.5629;
JHEP 1102 (2011) 067, arXiv:1012.3137

\bibitem{CP} L.Chekhov and K.Palamarchuk, 
Mod.Phys.Lett. {\bf A14} (1999) 2229-2244, hep-th/9811200

\bibitem{Cmamo} T.Morris, Nucl.Phys. {\bf B356} (1991) 703-728\\
Yu.Makeenko, Pis'ma v ZhETF, {\bf 52} (1990) 885

\bibitem{MMMM} Yu.Makeenko, A.Marshakov, A.Mironov, A.Morozov,
Nucl.Phys., {\bf B356} (1991) 574-628

\bibitem{Virmult}
A.Marshakov, A.Mironov and A.Morozov,
Mod. Phys. Lett. A7 (1992) 1345-1360, hep-th/9201010\\
Ch.Ahn and K.Shigemoto,
Phys.Lett. B285 (1992) 42-48, hep-th/9112057

\bibitem{GKMU} A.Mironov, A.Morozov, G.Semenoff, 
Int.J.Mod.Phys., {\bf A10} (1995) 2015

\bibitem{Kaz2} G.Segal, G.Wilson, 
Publ.I.H.E.S., {\bf 61} (1985) 5-65\\
M.Kazarian, arXiv:0809.3263

\bibitem{FKN2} M.Fukuma, H.Kawai, R.Nakayama,
Comm.Math.Phys. {\bf 143} (1992) 371-403



\end{thebibliography}
\end{document}